\documentclass[AMA,STIX1COL]{WileyNJD-v2}

\articletype{Article Type}%

\usepackage{amsmath}
\usepackage{graphicx}
\usepackage{subfig}
\usepackage[utf8]{inputenc}
\usepackage[bb=boondox]{mathalfa}
\usepackage{array}
\usepackage{ulem}

\newcommand\mvec[1]{\mathbf{#1}}
\newcommand\mmat[1]{\mathbb{#1}}
\newcommand\vecU{\mvec{U}} 
\newcommand\vecUd{\mvec{U}^{\delta}}

\newcommand\vecF{\mvec{F}}
\newcommand\vecS{\mvec{S}}
\newcommand\matH{\mmat{H}}

\newcommand\ppmatrix[1]{\begin{pmatrix} #1 \end{pmatrix}}

\usepackage{color, colortbl}
\usepackage{xcolor}

\newcommand{\revI}{\textcolor{red}}

\newcommand{\citer}{\text}
\newcommand{\farhaeus}{F\aa rh\ae us}

\begin{document}

\title{Reduced-order modeling of hemodynamics across macroscopic through mesoscopic circulation scales}

\author[1,3]{Olivier Adjoua}
\author[2]{St\'ephanie Pitre-Champagnat}
\author[3]{Didier Lucor*}

\authormark{Olivier Adjoua \textsc{et al}}

\address[1]{\orgdiv{ISCD}, \orgname{Sorbonne-Universit\'e}, \orgaddress{\state{Paris}, \country{France}}}
\address[2]{\orgdiv{Laboratoire IR4M}, \orgname{CNRS, Universit\'e Paris-Saclay}, \orgaddress{\state{Villejuif}, \country{France}}}
\address[3]{\orgdiv{LIMSI}, \orgname{CNRS, Universit\'e Paris-Saclay}, \orgaddress{\state{Orsay}, \country{France}}}

\corres{*Didier Lucor, \email{didier.lucor@limsi.fr}}

\presentaddress{LIMSI, Campus universitaire 508, Rue John von Neumann, F-91405 Orsay Cedex, France}

\abstract[Summary]{We propose a hemodynamic reduced-order model bridging macroscopic and mesoscopic blood flow circulation scales from arteries to capillaries.  {\it In silico} tree like {vascular} geometries, mathematically described by graphs, are synthetically generated by means of stochastic growth algorithms constrained by statistical morphological and topological principles. Scale-specific pruning gradation of the tree is then proposed {in order} to fit computational budget requirement. Different compliant structural models with respect to pressure loads are used depending on vessel walls thicknesses and structures, {which vary considerably from macroscopic to mesoscopic circulation scales.} Nonlinear rheological properties of blood are also included and microcirculation network responses are computed for different rheologies. {Numerical results are in very good agreement with available experimental measurements.} The computational model captures the dynamic transition between large- to small-scale flow pulsatility speeds and magnitudes and  wall shear stresses, which have wide-ranging physiological influences.}

\keywords{microcirculation, hemodynamics, multi-scale, reduced-order modeling, pulsatility, F\aa rhaeus Lindqvist effect}

\maketitle

\section{Introduction}
\label{sec:introduction}
Microcirculation plays a central role 
for many important physiological mechanisms of the circulatory system. 
It is the main site for exchanges between  blood and  tissues and also regulates and reorganizes the blood flow according to metabolic activity or the development of certain pathologies, e.g. neurodegenerative diseases, pulmonary arterial hypertension or tumors. 
Microvascular structure is highly complex but very well distributed across spatial scales, thanks to a fractal type of geometry organized in branching patterns. As mentioned  {by \citer{Secomb et al [\cite{Secomb_model2008}]}} ``The microcirculation represents an important `mesoscale' in physiological systems, functionally bridging higher and lower scales." Indeed the functional status of the microcirculation, { which is associated} from now on to the vascular mesoscale, is determined by complex processes taking place at the cellular/molecular micro- and nanoscale and thus strongly influences tissues and organs. Reciprocally, large macroscale systemic quantities such as blood pressure and flux balance inevitably affect the function of the microcirculation.\\
Despite tremendous progress, reliable {\it in vivo} measurements of microcirculation structure and function remain very difficult. Numerical models are  needed 
to establish quantitative relationships with phenomena occurring on these larger and smaller scales. {According to \citer{Secomb et al.} [\cite{Secomb_model2008}] and \citer{Boyer et al.} [\cite{Boyer_2018}]}, several coupled and nonlinear submodels must interact at different scales in order to account for nonlinear blood rheology, microvascular morphology and extravascular pressure gradient. A blood flow model interacting with a microvascular architecture through vessels structural mechanical models and subject to a scale-dependent blood rheology model must be considered. Ideally, a solute transport model of blood compounds and its interaction with an arteriolar vasomotion model should also be considered.
Nevertheless, for each of these components, it remains impossible to simulate phenomena across these spatial multi-scales by means of 3D models { presented by \citer{Quarteroni et al.} [\cite{QUARTERONI2016193}] }
One-dimensional reduced-order models (ROM) are commonly used to simulate blood flow regimes for which pulse waves propagate in large compliant arteries. They are increasingly being used to study physiological hemodynamic phenomena {referenced by \citer{Mynard \it et al.} [\cite{Mynard2015}]} and to assist in the validation or measurement methods. Most studies focus on  macroscales and investigate the impact of mechanical modeling and calibration of the arterial walls {[\cite{dumas2016}]} , efficiency and robustness of deterministic and stochastic numerical schemes {[\cite{boileau2015benchmark,brault2017,Lucor2018}]} and boundary conditions  {[\cite{olufsen2000numerical}]} or detailed anatomy {[\cite{Blanco2015}]}.\\
While computational models of very different complexities and fidelities have been developed to model circulation at smaller scales,[\cite{FANTINI2014202,Fedosov2010,Obrist2010,Possenti2019}] very few topologically/geometrically-detailed network-based attempts exist.[\cite{Stamatelos2014,Lorthois2011}] 
Moreover, their reliability and use are still hampered mainly because they do not encapsulate all of the necessary submodels and are often dissociated from those of the systemic circulation. 
Other reasons include disregard of structural compliance 
and non-stationary flows, which are detrimental to the study of the significance of pulsatility in microcirculation {revised by \citer{Pan et al.} [\cite{pan2014Pulsatile}] and \citer{Lee { et al}.} [\cite{lee2007}]}. Indeed, {several experimental studies have shown that the pressure pulse, generated by the heart, reaches the capillary or even beyond [\cite{Salotto1986}]. In fact, microvascular} pulsatility has wide-ranging physiological influences, e.g. from the stimulation of nitric {oxyde} release [\cite{Fisher2001}] regulating  the microvascular tone responsible for fine control of local blood flow to the control of inflammation response during cardiopulmonary bypass operations. Finally,  non-linear blood rheology is often over-simplified in those models. \\
Computational domains representing large-scale complete and  anatomically accurate microvasculature 
are very difficult to construct from biomedical images. Indeed, these highly distributed and space-filling structures bear a high degree of spatial heterogeneity in architecture 
{and topology.}
The authors then rely on the inherited framework of one-dimensional ROM  to simulate the blood flow, solved thanks to a discontinuous Galerkin (DG) method. In section (\ref{sec:methods}), they show how to extend its capability toward microcirculation.
For the computational domain, a method is proposed to generate {idealized yet} realistic vessel trees by means of stochastic algorithms. {Several truncation strategies are also investigated in order to efficiently prune the tree} 
and to provide calibrated boundary conditions in place of the missing branches.
{The vascular tree is mathematically described by a graph of blood vessels.}
Moreover, vessels arterial walls being {very} different in terms of their response to pressure loads {from macroscopic to mesoscopic circulation scales}, different structural models are adopted to address the specifics of the wall thickness-to-radius ratios.
Finally, {the authors} have accounted for some of the blood rheology effects through the parametric calculation of a dynamic macroscopic description of the {\it apparent } blood flow viscosity.
Some simulation results for a vasculature tree ranging from macro to mesoscale differing by more than two orders of magnitude {are also presented and discussed in sections (\ref{sec:results}) and (\ref{sec:discussion}), respectively}. {It is through the inclusion of these several important multiscale modeling components, unified under a common computational framework capable of simulating the {\it dynamic} effects generated by a reasonably large circulatory network (i.e. $\mathcal{O}(10^4)$ vessels) from large arteries to capillaries, that this work is original.}

\section{Materials and Methods}
\label{sec:methods}

\subsection{Microcirculation reduced-order model}
\label{subsec:MCmodel}
One-dimensional ROM are commonly used to simulate convection-dominated blood flow for which pulse waves propagate in large elastic arteries.[\cite{BOLLACHE2014424,boileau2015benchmark,brault2017}] They rely on a fluid-structure interaction mathematical framework that is much simplified when assumptions of Newtonian properties, linear elasticity {(and to some extent nonlinear viscoelasticity)} and homogeneous geometry are made for the blood fluid, the vessel walls mechanical response and the circulation network, respectively. They predict hemodynamic quantities
with satisfactory accuracy and their low computational cost enables the simulation of broad arterial networks.[\cite{Blanco2015}] {\revI authors of the present paper} rely on this inherited framework to enrich it and extend its capability toward microcirculation at mesoscopic scales both in terms of  modeling and computational efficiency.

\subsubsection{One-dimensional modeling framework}
\label{subsec:1Dmodel}

We consider an arterial tree of $N$ vessels, mathematically described by a graph: i.e. a network of thin, deformable, {and axisymmetric} arterial segments filled with blood, taken as an incompressible non-Newtonian fluid. The formulation directly derived from Navier-Stokes equations, for each straight arterial segment of lumen diameter $D$, based on the conservation of mass and momentum equations and a pressure law, takes the form of a nonlinear system:
\begin{eqnarray}
  \frac{\partial A}{\partial t} + \frac{\partial Au}{\partial x} & = & 0 \nonumber \\
  \frac{\partial u}{\partial t} + (2\alpha-1)u \frac{\partial u}{\partial x} + (\alpha-1)\frac{u^2}{A} \frac{\partial A}{\partial x} & = & - \frac{1}{\rho} \frac{\partial p}{\partial x} + \frac{f}{\rho \,A} \nonumber \\
  p & = & p_{\text{ref}} + \psi \big (A,A_{\text{ref}},\beta \big ),
\label{eq:govEq}
\end{eqnarray}
where $t$ denotes time, $x$ is the axial coordinate, $A(x,t)=\pi D(x,t)^2/4$ is the circular cross-sectional lumen area, $u(x,t)$ and $p(x,t)$ are cross-sectional averaged velocity and internal pressure, $\rho$ the blood density, $f$ the friction force 
related to the choice of axisymmetric velocity profile through $\displaystyle f = -2\mu \pi \frac{\alpha}{\alpha -1}u$, $\mu$ is the apparent blood dynamic viscosity, $\displaystyle \alpha=\int_0^R 2r\, \Tilde{u}^2 / (uR)^2 dr$ is the momentum flux correction (Coriolis) coefficient accounting for the nonlinear integration of radial velocities in each cross-section
~[\cite{QUARTERONI2016193}], and $\Tilde{u}=u\,\phi(r)$, with the velocity profile shape $\phi(r)=\frac{\alpha}{2-\alpha}\left (1-\left ( \frac{r}{R}\right )^{\frac{2-\alpha}{\alpha-1}} \right )$.\\
The ``tube law'' relation, links the blood pressure to the vessel wall deformation, through a chosen function $\psi$, satisfying $(\frac{\partial \psi}{\partial A}>0,\; \psi(A_{\text{ref}})=0)$, which may reflects compliant effects through some elastic model, cf. subsection (\ref{subsubsec:wall}). The underscript $\cdot_{\text{ref}}$ denotes quantities in a reference {\it unloaded} state, $A_{\text{ref}}$ 
and $p_{\text{ref}}$ are taken both {\it constant} along each artery.\\ 
Due to the interaction between the pulsatile blood pressure and the vessels distensibility $\displaystyle \mathcal{D}=\frac{1}{A} \frac{dA}{dp}$, flow/pressure waves travel at a certain speed $c(x,t)$ in the system. Wave propagation and attenuation strongly depend on $\mathcal{D}$ and the ratio between fluid pulsation and viscous forces through the Womersley number: $W_{\alpha} = \frac{D}{2}\left (\frac{\rho \omega}{\mu} \right )^{1/2}$, proportional to the oscillation angular frequency $\omega$.\\ It is worthwhile reminding the main results obtained for a {\it linearized} version of {Equation \ref{eq:govEq}} {with purely elastic arterial wall and viscous fluid} under different regimes. The approach considers small perturbations and looks for solutions of the form:  $\big (u,A,p \big )(\omega,x,t)=\big (\hat{u},\hat{A},\hat{p}\big ) (\omega,x=0)\exp{(i(\omega t -kx))}$ with  $k=k_r+i\, k_i$, the complex wave number with $c=\omega/k_r$ the wave speed and $\gamma(\omega)=-2\pi k_i/k_r$ the attenuation coefficient.[\cite{Van2Vosse_LN}] Different scenario take place depending on the value of $W_{\alpha}$:
\begin{itemize}
    \item[--] 
for large $W_{\alpha}\rightarrow \infty$, the viscous term is neglected and simplification of the system leads to a pressure wave equation. The wave speed  $c(w)=c_{\text{ref}}=\sqrt{1/\rho \mathcal{D}_{\text{ref}}}$ is constant with no wave attenuation: $\gamma(\omega)=0$. Concomitant forward/backward pressure waves travel  at constant speed without damping. 
\item[--] 
for small $W_{\alpha}\rightarrow 0$,   inertial  term is neglected and the viscous term is modeled by a Poiseuille law. This simplification leads to a pressure diffusion equation. The wave number is complex with: $c(w)=\frac{1}{2} W_{\alpha} c_{\text{ref}}$ and $\gamma(\omega)=2\pi$.
\item[--] 
for arbitrary $W_{\alpha}$, the wave number is:
\begin{equation}
    k(\omega)=\pm \frac{\omega}{c_{\text{ref}}}\left ( \frac{1}{1-F_{10}(\omega)} \right )^{1/2}, \;\, \text{with} \;\;F_{10}(\omega)=\frac{2J_1(i^{3/2} \,W_{\alpha})}{i^{3/2} \,W_{\alpha}\,J_0(i^{3/2} \,W_{\alpha})},
    \label{eq:wavespeedWomersley}
\end{equation}
from which wave speed and attenuation are readily computable ($J_{0,1}$ are Bessel functions).
\end{itemize}
The computational model should be able to accurately capture all different pulse waves interacting at different speeds through the system.
The construction of the computational domain, blood rheology model and vessel wall structural models will be described next. Let us start with the  generation of a healthy vascular network.

\subsubsection{Vascular tree generation and graph models}
\label{subsec:graph}
Microcirculation models of healthy tissues are usually established based on physical microvascular network structures,[\cite{pan2014Pulsatile,lee2007}] which, by nature, bear some degree of spatial heterogeneity in architecture (vessels diameters and lengths, vascular density) and topology (hierarchical organization, deviation from symmetry). Relative to normal tissues, tumoral tissues microcirculation (not treated here) exhibits higher structural and functional heterogeneity.
With great effort, microvascular networks may be reconstructed from measurements (e.g. by intravital microscopy). 
The vasculature is organized in different patterns depending on the scale under consideration, e.g. arteries make a divergent arborescence, while the capillaries form some kind of ``mesh-like" space-filling network.[\cite{Lorthois2010}] \\
Another approach is to generate vessel trees {\it ex novo} by means of mathematical algorithms, across all scales down to the capillaries size, (statistically) constrained by some realistic morphological and topological principles, e.g. table (\ref{tab:vascular_structure}).[\cite{Takahashi2014,Van2Vosse_LN}]

\begin{table}[hbt]
\caption{Representative vessels distribution and size in the human circulatory system.[\cite{phdArimon,Blanco2015}] Here the number $n$ is typical of the chosen averaged diameter $D$ presented in the table. Vessels categories may as well be thought in terms of size ranges; for instance, small arteries in $[600-100]\mu m$, arterioles in $[100-10]\mu m$ and capillaries in $[10-5]\mu m$.}
\label{tab:vascular_structure}
\centering
\scriptsize
	\centering
	\begin{tabular}{|l|c|c|c|c|c|c|c|}
	\hline
		& \multicolumn{3}{c |}{{macroscale}} & \multicolumn{3}{c |}{{mesoscale}} & {macroscale} \\
		\hline
\hline
	& Aorta & \multicolumn{2}{c |}{Arteries} & Arterioles & Capillaries & Venules & Veins \\
 \hline
 &  & large & small &  &  &  &   \\\hline
Number $n$  &  1	& 40 & 7100 & $1.6\cdot 10^8$ & ${5.5\cdot 10^9}$ & ${10^{9}}$ & ${10^{2}}$ \\
Lumen diameter $D$ & $2.6\,cm$ & $8\, mm$ & $600\, \mu m$ & $20\, \mu m$ & $9\, \mu m$ & $20\, \mu m$ & $5\, mm$ \\
Thickness $h(\%D$) & $5.8\%$ & $8.2\%$  & $18.6\%$  & $32.5\%$ & $10\%$ & $\approx 10\%$ & $\approx 10\%$ \\
\hline
\end{tabular}
\end{table}

\paragraph{Some notations and concepts}
We now consider a circulatory tree $\mathcal{T}$ organized into small arteries,
arterioles and capillaries (draining venules and veins are voluntarily not included and replaced by appropriate boundary conditions). The tree   is split
into $N$ vessels $v_{.}$, such that $\mathcal{T} = \bigcup_{n=1}^{N} v_n$ . Each vessel $v_n$ has its own constant length $l_n$ and reference lumen diameter $D_{n,\text{ref}}$ (we will often omit the $n$ subscript for convenience). Junctions between vessels are simply modeled as nodal points.\\
Biological branching systems most often exhibit fractal nature, i.e., a scale-independent self-similarity in the bifurcation pattern of their architecture. The measured fractal dimensions of vascular systems reveal a power function with a single exponent $<\cdot>^{-d_{\text{f}}}$ expression. This fractal dimension has been reported to be $d_{\text{f}}\in[1.1,1.86]$[\cite{Lorthois2010}]. On the basis of a stochastic model relating the probability of branch density with aggregated branch length, morphological parameters such as surface area or content volume may be expressed through definite integrals of $d_{\text{f}}$ for a large range of diameters.[\cite{Takahashi2007}]
Here, diameter distribution is based on the law of Murray, which 
relates parent and daughter vessels at bifurcations as in {\citer{Murray {\it et al}.} [\cite{murray1926physiological}]}:
\begin{equation}
D_p^m = D_{d1}^m  + D_{d2}^m,
\label{eq:murray_law}
\end{equation}
with $m\approx 3$\footnote{{Reported measured values in various vascular beds give} $m\in[2.7,3]$ [\cite{Takahashi2014}]}. {This universal law [\cite{Sherman431}] was proposed for the optimal design of blood vessels, based on a balance between blood circulation and metabolic powers. While this law holds true for mid-scale circulation, this is not the case for the larger vessels, where inertia and turbulence effects may occur and for microcirculation due to shear-thinning effects. When F\aa hreus-like phase separation effects occur in the latter case, thanks to some simplified blood constitutive model (e.g. Qu\'emada's model), it is possible to derive an extended version of Murray’s law, and see how it affects the optimal geometries of fractal trees to mimic an idealized arterial network. In this case, it  was shown that the geometric scaling becomes less obvious as the optimal factors are now dependent on the generation index, the size of the root vessel and the relation between vessel diameter and mean shear rate [\cite{Moreau2015}]. In the following, a simpler hands-on approach is chosen, for which the Murray's law is calibrated differently depending on the vascular scale.}

For a given parent diameter $D_p$, a relation of proportionality $D_{d1}=\kappa \, D_p$ is assumed, and $D_{d2}$ is obtained from Equation \ref{eq:murray_law}. The branch length-diameter relationship follows a power law: $l_{n}=\zeta \times D_{n,\text{ref}}^{\gamma}$,[\cite{Takahashi2014}] with $m=d_{\text{f}}+\gamma$, and $\zeta$ a scaling determined according to other relations, cf. \citer{Takahashi}.[\cite{Takahashi2014}]
~\\ In our case, the coefficients $\kappa$ and $(m,\zeta,\gamma)$ will be random and deterministic, respectively. Their respective adjustments will depend on the vascular type and domain as explained next.
{As derived in \citer{Kamiya et al.}[\cite{Takahashi2007}], it is possible to relate the quantity $\zeta$ to the knowledge of the fractal dimension, the junction exponent, geometrical bounds of largest $R_{max}$ and smallest $R_{min}$ vessel radius in a fractal tree and its approximate volume $V$: $\zeta = \lambda\iota /R_{max}^{d_\mathrm{f}+\gamma} $ where 
$ \lambda=V(-d_\mathrm{f}+2)/(\pi\iota(R_{max}^{-d_\mathrm{f}+2}-R_{min}^{-d_\mathrm{f}+2}))$ with
$\iota = d_f/(R_{\mathrm{min}}^{-d_\mathrm{f}} - R_{max}^{-d_\mathrm{f}}) $. The entire adult human arterial volume approximate $V\approx 700 \mathrm{ml}$ from aorta down to capillaries [\cite{Takahashi2007} ]. Data provided by \citer{Boileau et al.}[\cite{boileau2015benchmark}] allow us to frame the volume of the macrocirculation tree. Under the assumption of a fractal tree, the corresponding $\zeta$ parameter can be found thanks to the model. The same procedure is then used to evaluate the microcirculation tree length parameter. }

\paragraph{Graph representation and Stochastic growth model}
For computational purpose, the topological network is  reduced  to a graph, described by  node coordinates and a connectivity matrix. The graph is a full binary tree, which every node has either zero or two children. Its representation is very handy as it naturally provides (statistical) information related to the connectivity and the geometry of the entire (or part of the) network. 
Our tree generation is inspired by stochastic growth models (e.g. RSB model[\cite{Werner1973}]). It starts from a root vessel ($D_{\text{root}}$) and adds new segments at each generation in an iterative stochastic process, until a stopping criteria is met (e.g. minimal capillary size). After numerous trials, our search of realistic trees, provided us with robust values of $(\kappa,m,\zeta,\gamma)$ as described in table (\ref{tab:random_growth}).

\begin{table}[hbt]
\caption{Numerical values used at each network bifurcation, during the stochastic generation of the vascular tree, to determine vessels  diameter and length. Given a parent vessel of diameter $D_p$ at a certain generation, daughter vessels diameters are obtained as:  $D_{d1}=\kappa \, D_p$, with $\kappa \sim \mathcal{U}_{[\kappa_{\text{min}},\kappa_{\text{max}}]}$ and $D_{d2}=(1-\kappa^m)^{\frac{1}{m}} \, D_p$, according to Murray's power law. Concerning the branch length, except for the capillaries, the following deterministic relation with the diameter is used [\cite{Takahashi2007}]: $l_{.}(D_{.})=\zeta (D_{.})^{\gamma}$ with $m=d_{\text{f}}+\gamma$, and $\zeta$ determined according to the relations and measurements introduced in \citer{Kamiya \it et al.}[\cite{Takahashi2007}]. A fractal dimension of $d_{\text{f}}=1.75$ is considered in this work. }
\label{tab:random_growth}
\centering
\scriptsize
	\centering
	\begin{tabular}{|l|c|c|c|c|c|c|}
\hline
 $D_p$ range &  $\kappa_{\text{min}} (\%)$ & $\kappa_{\text{max}} (\%)$ & $m$       & $d_{\text{f}}$ &  $\zeta$ & $\gamma$ \\
 \hline
$D_p \in [2.6cm,500\mu m]$   &  50	& 98 & 2.7 & 1.75 & 104 & 0.95  \\
 $D_p \in [500\mu m,100\mu m]$ & 80 & 95 & 2.7 & 1.75 & 1453 & 0.95  \\
$D_p \in [100\mu m,10\mu m]$ & 65 & 79  & 3 & 1.75  & 1453 & 1.25  \\
$D_p < 10\mu m$ & 75 & 95  & 3.5 & n/a & n/a & n/a \\
\hline
\end{tabular}
\end{table}

\begin{figure}[ht]
    \begin{center}
\includegraphics[width=\textwidth]{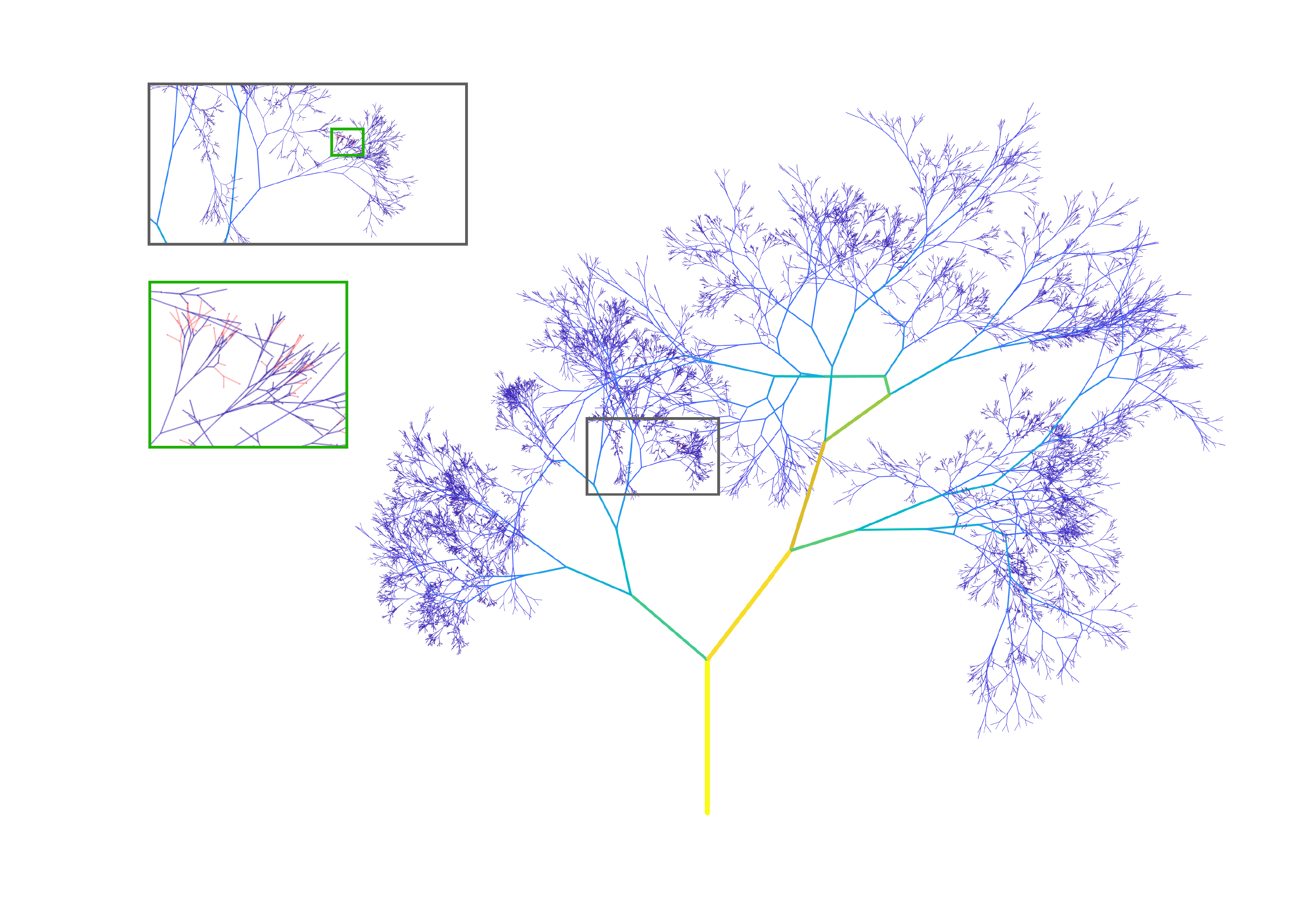}
\caption{Example of a complete network graph with $N=4015484$ segments, generated from $D_{\text{root}}=1mm$ and $D_{\text{min}}=7\mu m$, with a total of $6$ medium arteries, $1294$ small arteries, 1200309 arterioles and 2813875 capillaries; for a total length of $2396.44m$ and volume $0.9708\times 10^{-6}m^3$. The pseudo-3D space visual representation of the 1D graph is arbitrary. Moreover, only the fifteen first generation segments are represented on the figure, but the connectivity, length and diameter scales of the depicted segments are up to scale. The colors follow the diameter sizes.}
\label{fig:network_big}
    \end{center}
\end{figure}

\begin{figure}[ht]
\begin{center}
\includegraphics[width=0.4\textwidth]{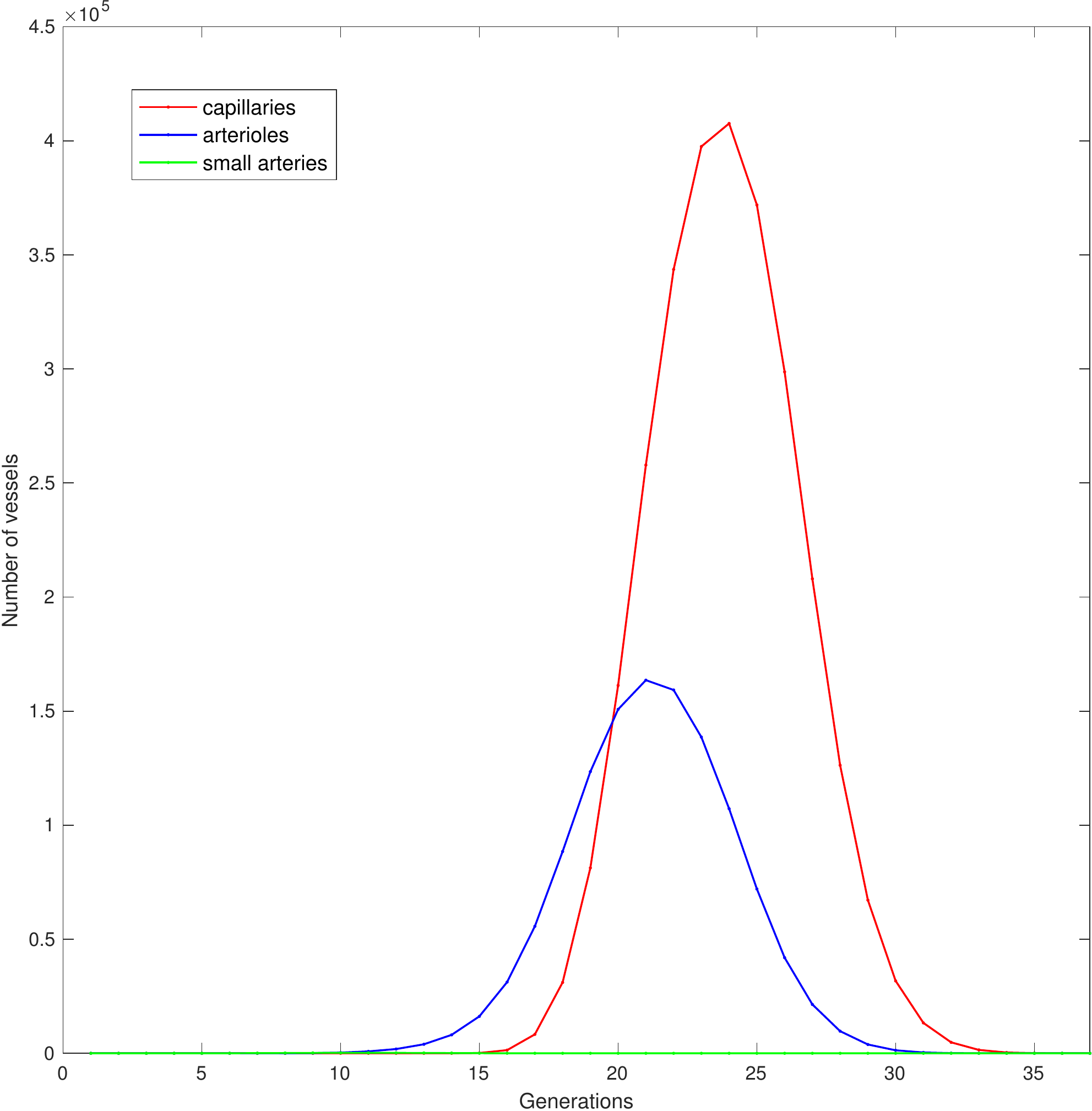}(a)\hspace*{5mm}
\includegraphics[width=0.4\textwidth]{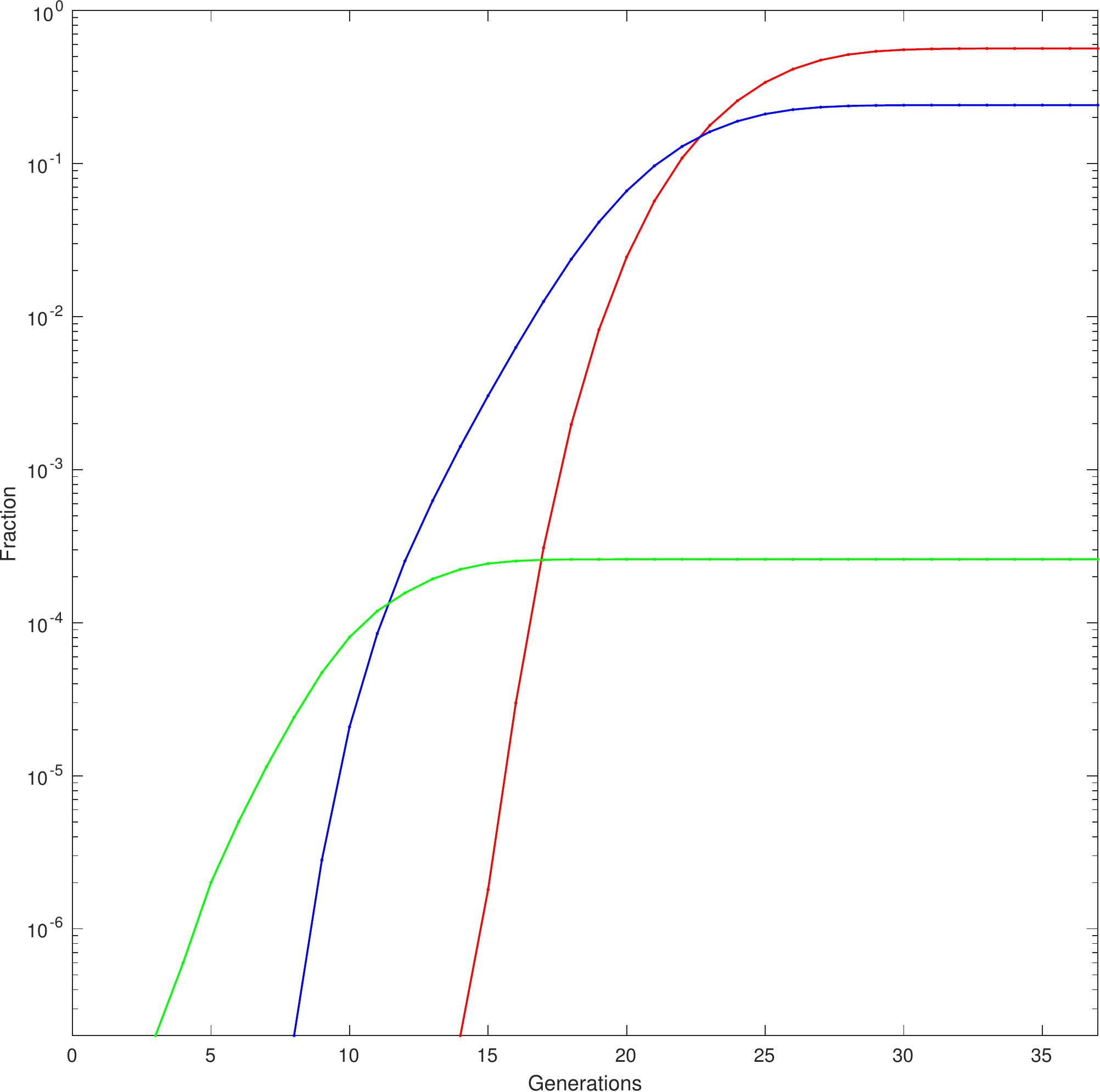}(b)\\
\vspace*{5mm}
\includegraphics[width=0.4\textwidth]{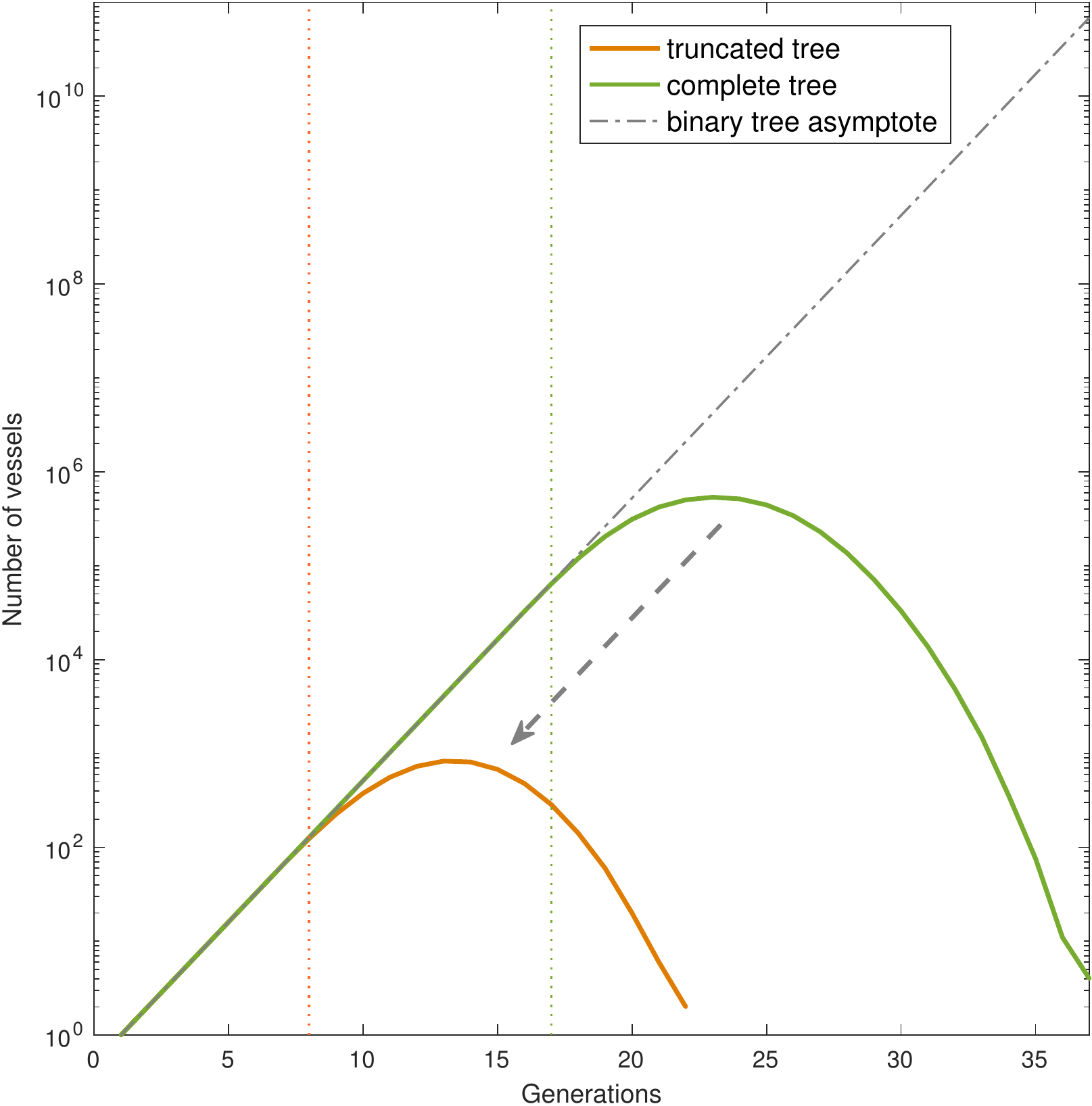}(c)\hspace*{4mm}
\includegraphics[width=0.43\textwidth, viewport=210 0 720 485, clip]{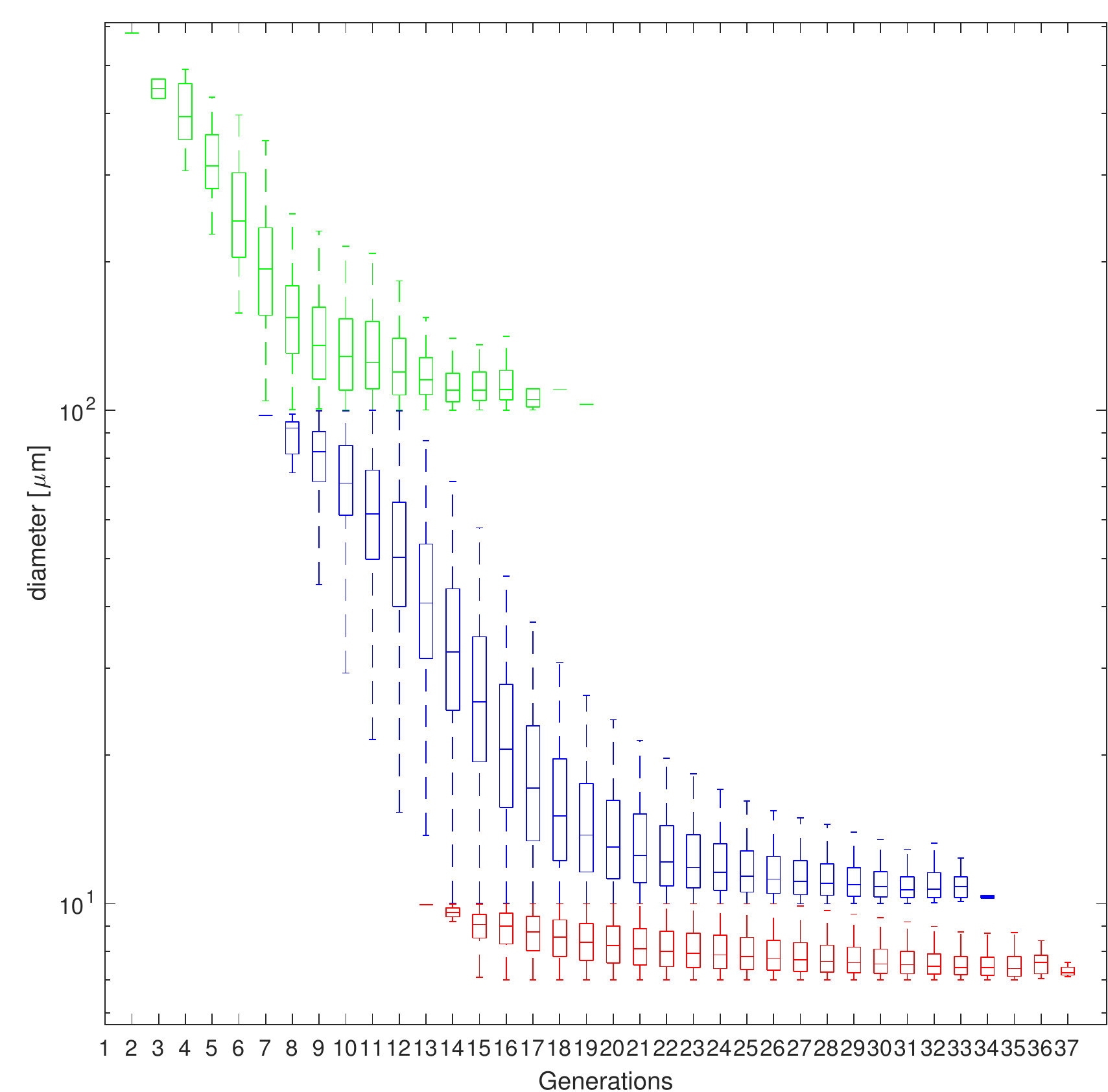}(d)
\caption{Distributions of small artery (green), arteriole (blue) and capillary (red) generation numbers (a), total vessels number (c), their fractions (b), diameter size statistics, of Box-plot type, (d) relative to all segments of a given generation level for the microvasculature illustrated in Figure~\ref{fig:network_big}.}
\label{fig:network_statistiques}
\end{center}
\end{figure}

The bounds of the $\kappa$ probability measure are adjusted in order to provide vessels distributions, whose statistics approach our reference, cf. table (\ref{tab:vascular_structure}). {We notice that the $\kappa$ range for the arterioles is consistent with the optimal prediction of Moreau \& Mauroy [\cite{Moreau2015}].} We note that the Murray's exponent has to be slightly adjusted for arteries and arterioles. 
While the same $d_{\mathrm{f}}$ is considered for the entire arterial tree, the calibration of $\zeta$ is made on two different scales, i.e.:  on a fractal tree with extremum diameters $(12mm,500\mu m)$ and volume $200ml$,[\cite{Takahashi2007,phdArimon}] and on a more distal tree with diameters $(600\mu m,8\mu m)$ and volume of $0.043 ml$.[\cite{Takahashi2007}]\\
Concerning the capillaries we still generate them with a tree structure but, as the capillary bed is often made of a  random homogeneous mesh-like structure, we have to increase the exponent to a larger value ($m=3.5$), in order to produce {a larger number of} vessels, {in agreement with physiological studies [\cite{silbernagl2001atlas}]}. Moreover, we rely on a random distribution to generate their lengths $l_{\text{cap}} \sim \mathcal{U}_{[500\mu m,700 \mu m]}$ according to the literature.\\
{Figure~\ref{fig:network_big}} shows an example of a generated network   ranging from macroscopic ($D_{\text{root,ref}}=1mm$) to mesoscopic ($D_{\text{min,ref}}=7\mu m$) scale. {The feeding vessel, despite being of moderate size (i.e. $1mm$), does represent the macroscale. Without restriction, it is possible to start with a larger vessel (e.g. $\mathcal{O}(cm)$), by connecting to the current model a network of large arteries such as the 55/56-artery networks frequently used in the literature, [\cite{Alastruey_2008,boileau2015benchmark,brault2017}]}.
Not all vessels are represented {in the figure}, but the connectivity, length and diameter of the depicted segments are up to scale.  Zoomed-up inset plots reveal some fine details (capillaries represented as red vessels). Figure~\ref{fig:network_statistiques} summarizes different geometric statistical markers along the different generations of that network, differentiating among small arteries ($D>100\mu m$, {in green}), arterioles ($10\mu m<D<100\mu m$, {in blue}) and capillaries ($D<10\mu m$, {in red}). The distributions in Figures~{\ref{fig:network_statistiques}(a)-\ref{fig:network_statistiques}(b),\ref{fig:network_statistiques}(d)} exhibit very similar features as the ones described by imaging capabilities{, such as:} the spread of the capillaries across $\sim 20$ generations{, a fraction of capillaries close to $60\%$, a hyperbolic decline of averaged arteriolar diameter with vessel generation,} as in [\cite{prieschapter1_2011}]). The topological heterogeneity of the network is also well captured in the statistics of Figure~\ref{fig:network_statistiques}(d), with very different diameters (and lengths, not represented) within a given generation. {Skewed distributions (with positive skewness) are obtained for small arterioles (here after the 17$^{\text{th}}$ generation), in agreement with the literature [\cite{prieschapter1_2011}]. These types of analysis were repeated several times for the generation of multiple networks. By doing so, it was checked that} the proposed algorithm produced statistically robust and coherent graphs. 

\begin{figure}[ht]
    \begin{center}
    \includegraphics[width=0.45\textwidth]{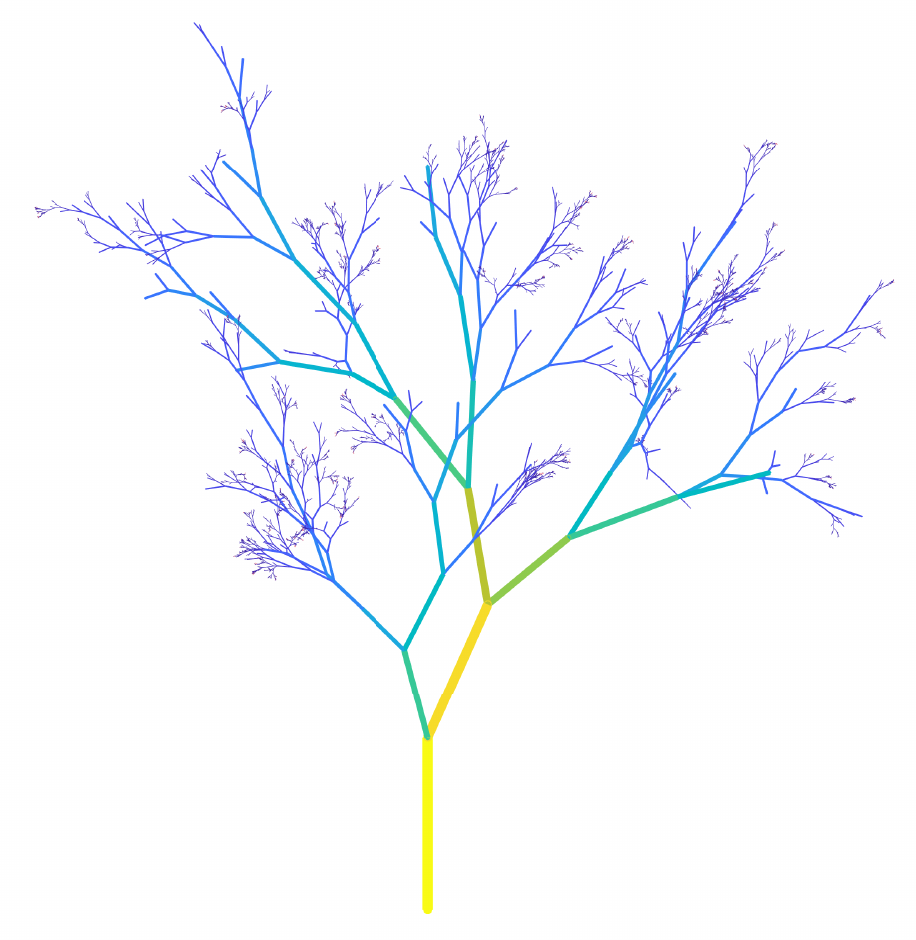}
    \caption{Truncated network graph ({pseudo-}3D view) with $N=5301$ segments and 2649 bifurcation nodes generated from $D_{\text{root}}=1mm$ and $D_{\text{min}}=8.5\mu m$, with a total of $435$ small arteries, 4548 arterioles and {312} capillaries; for a total length of ${12.57m}$ and volume $454ml$.
    }
\label{fig:network_small}
    \end{center}
\end{figure}

\paragraph{Truncation strategy}
Hemodynamics simulations of such large-scale complex networks as in Figure~ \ref{fig:network_big} are obviously out of computational reach. We must propose a proper way of truncating the network and provide calibrated boundary conditions in place of the missing branches {such that the simulation in the reduced domain produces a sound response.}
We have designed and tested several methods to prune the network based on deterministic or probabilistic trimming around one (or more) typical diameter(s) or generation scale(s) or based on imposing a decaying sequence of vessel growth inspired by the complete tree structure. 
The former approaches resulted in quite irregular distributions of vessels across generations. 
{Moreover algorithmic probabilistic ingredients seemed essential in the selection process in order to avoid strong clustering. But these model reduction approaches remain useful if one is interested in keeping the full description of a particular region of the vasculature, while sacrificing large remaining domains through early-generation trimming.}\\ 
The latter approaches gave best results. They rely on constructions that mimic the growth distribution of the complete graph. Figure~\ref{fig:network_statistiques}(c) shows how the binary-type growth of the complete tree (green solid line) deviates from the exponential asymptote {(oblique gray line)} close to the seventeenth generation  {(cf. vertical green dashed line)} due to the apparition of the capillaries, before to saturate and decline.  This sequence is rescaled according to the targeted computational budget in order to keep a similar smooth profile (orange solid line) and probabilistically imposed to produce a more regular trimming.  {The new distribution departs earlier from the asymptote due to the imposed trimming. Nevertheless, the right tail of the distribution persists across a generation range over which capillaries take birth, insuring the presence of those (although in smaller number) in the truncated network. } \\
Figure~\ref{fig:network_small} shows an example of a truncated network generated according to that pathway and ranging from macroscopic ($D_\text{root,ref}=1mm$) to mesoscopic ($D_{\text{min,ref}}=8.5\mu m$) scale. There are 312 capillaries (not visible) quite uniformly distributed within the network. 

We have now to include the blood rheological effects into the model. 

\subsubsection{Blood rheology}
\label{subsec:rheology}
{Low shear regimes exist in the blood circulation due to pulsatile, recirculating or bifurcating effects, making blood a shear-thinning, viscoelastic fluid with non-Newtonian rheology. Aggregation and disaggregation of red blood cells (RBC) in elongated «rouleaux» structures as well as their elastic properties and related dissipation mechanisms are mainly the cause of this complex behavior which has significant influence on various healthy and pathological physiological processes, particularly at microcirculation scale. Most accurate time-dependent non-Newtonian constitutive models include viscoelastic, thixotropic or shear-thinning effects (e.g. Maxwell model [\cite{Owens2008}] or Oldroyd-B model [\cite{Anand2013}]). Nevertheless, validation in large-scale arterial networks with vessel wall compliant effects are often neglected and only very few works are available for 1D applications [\cite{GHIGO2018}].}\\
{In this work, we follow a simpler approach accounting for non-Newtonian effects through a viscous contribution. Indeed, as} blood is made of a suspension of solid particles in a plasma fluid, 
it makes non-constant its {\it apparent} (or effective)  viscosity. The volume fraction of solid blood components named hematocrit $H$ is almost entirely made of RBC. It is common to distinguish among instantaneous tube hematocrit $H_t$ and {hematocrit flux passing though a point in a given  time as}
discharge hematocrit flux, $H_d$.
An interesting phenomenon appears in vessels of small size, e.g. $D<300\mu m$: a plasma layer (PL) forms at the lumen wall while RBC tend to concentrate closer to the vessel center. This heterogeneity is in fact at the source of complex nonlinear rheological blood properties in microcirculation. The literature  highlights three main effects.[\cite{Popel_ARFM2005, geddes2010bloodflow}] \newline
Given a certain $H_d$, numerous studies notice a decreasing trend in the apparent blood viscosity with decreasing vessel diameter  ($D>20\mu m$) where it starts increasing significantly ($D\leq 20\mu m$). This causes the change in the apparent viscosity of a Poiseuille blood flow and is the so-called {\it \farhaeus-Lindqvist} effect.
\newline
The {\it plasma skimming} effect describes {the difference between tube and discharge hematocrit, and therefore}
the asymmetric distribution of hematocrit in channels downstream a bifurcation.
[\citep{pries1989red}] Indeed, hematocrit transported near a junction, meet daughter branches of different sizes and flow rates. Considering the hematocrit and Poiseuille cross-sectional profiles, the distribution may become unequal.
This effect is more significant for (feeding) vessels of diameter $<40\mu m$.  \newline
Finally, another effect results from the velocity difference between $H_t$ and PL within the vessels and how it affects the local blood flow. As $H_t$ velocity is much higher than PL velocity, the hematocrit profile varies along the vessel,  
inducing longitudinal variations of the PL speed. This is known as the {\it \farhaeus} effect.

\paragraph{\farhaeus-Lindqvist effect}
In the literature, numerous {\it in vitro/vivo} works have collected and fitted  their data to parametric empirical descriptions of the apparent blood viscosity. For {\it in vivo} data, they have proposed a {phenomenological} relation for the relative apparent viscosity $\eta$, ratio between the blood and  plasma viscosities. {Derivations details for these models, provided here, are given in these references} [\cite{pries1992blood,Pries1994}]:
\begin{equation}
\eta = \left[1+(\eta_{0.45}-1)\cdot\frac{(1-H_d)^C-1}{(1-0.45)^C-1}\cdot\frac{D}{D-1.1}^2\right]\cdot\left(\frac{D}{D-1.1}\right)^2,
\label{eq:farhaeus_lindqvist_in_vivo}
\end{equation}
with:
\begin{equation}
C =\left(0.8+\exp(-0.0075D)\right)\cdot\left(-1+\frac{1}{1+D^{12}\cdot 10^{-11}}\right)+\frac{1}{1+D^{12}\cdot 10^{-11}},
\label{eq:farhaeus_lindqvist_C_shape}
\end{equation}
and:
\begin{equation}
\eta_{0.45}=6\exp(-0.085D)+3.2-2.44\exp(-0.06D^{0.0645}),
\label{eq:hematocrit045_in_vivo}
\end{equation}
the nominal apparent viscosity
for $H_d=45\%$.
\noindent Therefore, in this work we account for blood rheology effects through a dynamic macroscopic description of an apparent viscosity, which impacts the friction term in the momentum equation balance \ref{eq:govEq}:
\begin{equation}
f = -2\pi\underbrace{\mu_{\text{plasma}}\,\eta\big (D(A),H_d\big)}_{\mu}\frac{\alpha}{\alpha-1}u,
\label{eq:def_f_visc}
\end{equation}
with $\mu_{\text{plasma}}$ taken as the water viscosity and $D$ in $\mu m$. We have implemented and successfully tested this model against some literature benchmarks [\cite{lee2007}]. We have verified that accounting for the \farhaeus-Lindqvist effect induced an increased systolic flow by 10 to 45\% when reaching $30\mu m$ diameter arterioles, while decreasing it by 10 to 30\% in capillaries, compared to Newtonian simulations. 
Those results supported the choice of considering such effect for $D<300\mu m$.

\paragraph{\farhaeus ~and plasma skimming effects}
The use of a transport equation is a simple attempt to model (part of) the local \farhaeus effect. In this case, $H$ is passively transported at the flow velocity and retro-acts on the system through the blood viscosity, cf. Equation \ref{eq:def_f_visc}. We have implemented and validated this approach within the DG framework (cf. \ref{subsubsec:scheme}) for stationary blood flow velocity in a compliant vessel. Nevertheless, as mentioned in, [\cite{phdAudebert}] for pulsatile or even reverse time-dependent flow, we have noticed some spurious compression effects arising in the hematocrit concentration. 
These oscillations were not only detrimental to the numerical simulation (unstable situation) but they also provided very different results compared to the much smoother observed concentrations for experimental injection of a passive compound in arterial circulatory subsystem. 
Due to these difficulties, we have decided not to include the transport equation in this study, to the price of disregarding the plasma skimming effect.\\

\subsubsection{Vessel wall structural model}
\label{subsubsec:wall}

The vessel walls of arteries and arterioles are composed of distinct layers. For instance elastic fibers included in the tunica media of large arteries provide vessels with an elastic component. Dynamic changes in the vessels geometry reflect both active and passive regulation mechanisms. 
Transmural pressure $\Delta p_t$ 
passively  affects the lumen diameter. Ignoring the shear and inertial components, this effect may be modeled thanks to a Laplace's law, 
leading to an algebraic relation. This simple {\it thin} membrane law  has been successfully exploited in many applications, 
but is not well adapted for arteriolar wall bearing a thick muscularis layer designed to sustain high transmural pressures. In the following, we adopt the structural models described in [\cite{causin2017mathematical}] in order to address the specifics of the wall thickness-to-radius ratios. \\
Each vessel segment is modeled as an elastic but incompressible cylindrical shell  with its own rigidity and only moderate deformations
are considered. 
Moreover, due to the lack of experimental data from the literature, Young modulus $E$ is considered constant for each vessel, through the range of transmural pressure. 
In practice, given a reference vessel diameter, $E$ is {piecewise linearly} interpolated from literature data.[\citep{causin2017mathematical}]
{It was checked that other types of interpolant (nonlinear polynomials, spline, shape-preserving piecewise cubic) did not significantly change the value of the Young modulus within the considered range (i.e. $\ell_2$-norm difference $\leq 7\%$).}

\begin{figure}[ht]
\begin{center}
\includegraphics[width=8cm]{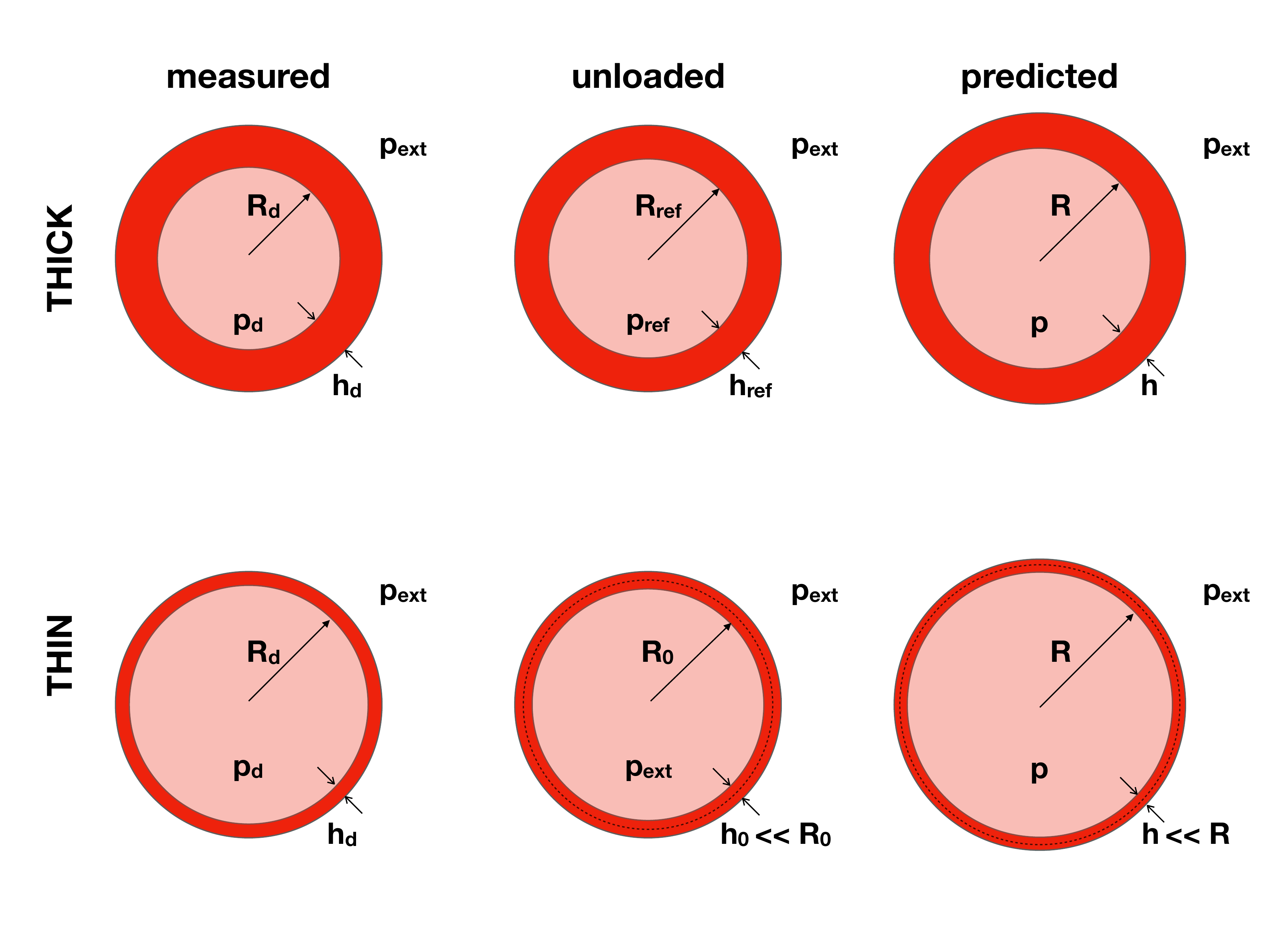}(a)\includegraphics[width=7.5cm]{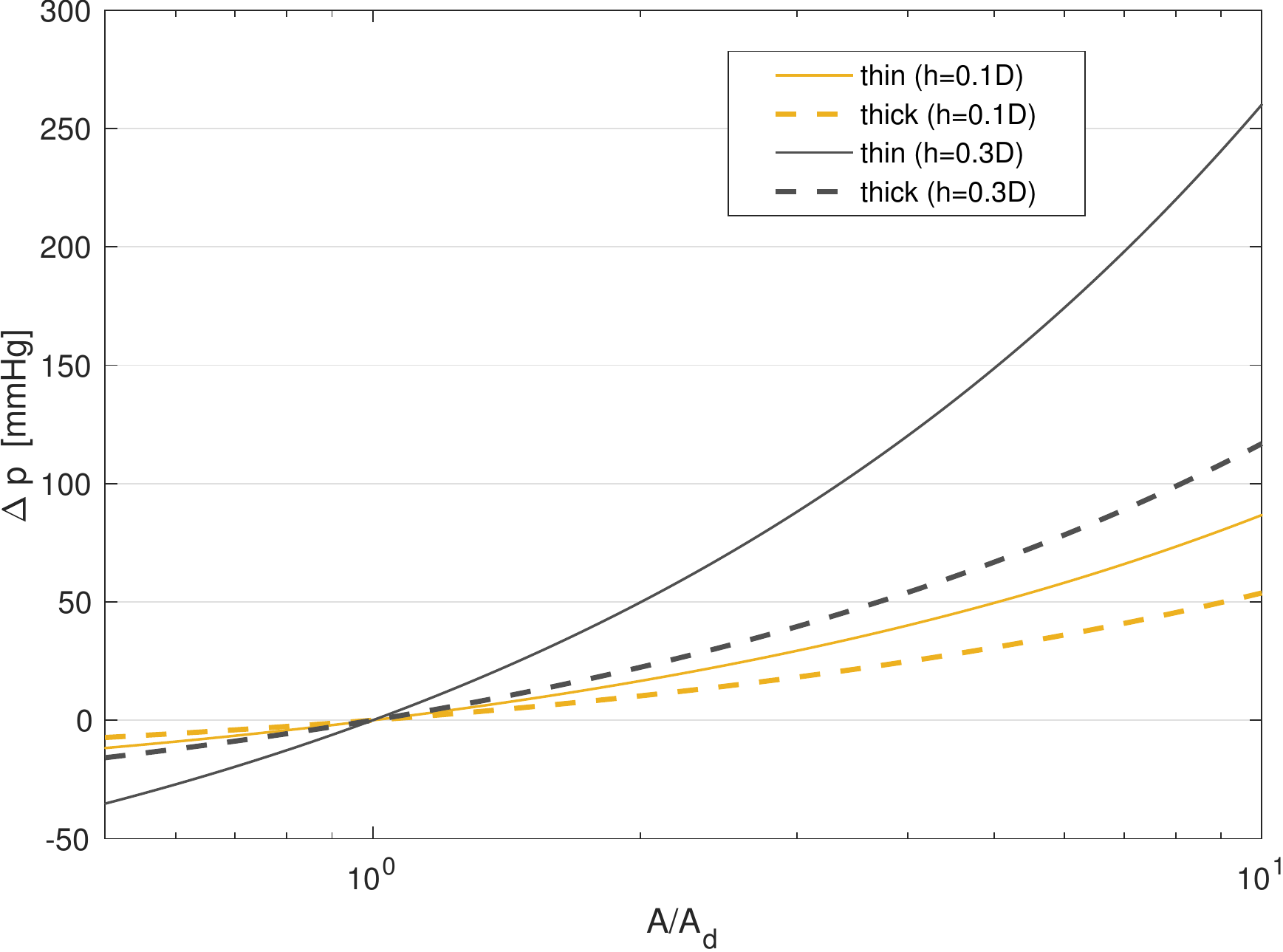}(b)
\caption{Structural model for arterial walls, (a): different cross-section configurations depending on the modeled wall thickness (thick model: top vs. thin model: bottom row). First column: measured geometries (when available); center column: reference unloaded configurations, used to provide the predicted response: right column.\\
(b): transmural pressure vs. cross-section area variation provided for two different initial vessel wall thickness $h$; $D_{\text{root}}=20\mu m$ and $E=20kPa$. Given a certain level of swelling of the lumen, the thick-walled model sustains lower transmural pressure than the thin-walled model. The thin-walled model provides unphysical response for thick unloaded vessel.}
\label{fig:generic_configuration2}
\end{center}
\end{figure}

In Figure~\ref{fig:generic_configuration2},
we report the notation required for the definition of
the relevant configurations. We refer to  $(R_{\text{ref}},h_{\text{ref}})$ to describe its reference undeformed condition. 
First are the experimentally measured {\it in vivo} geometries, then comes the unloaded
configurations: stress-free geometry for thick-walled rings and, zero
transmural pressure geometry for thin-walled rings, respectively. Finally, predicted geometries are described. 

~\\
The vessel wall motion induced by the  transmural pressure is described by a  linear elasticity relation, based on the reference {\it unloaded} state:
\begin{equation}
p = p_{\text{ref}} + \beta(\sqrt{A} - \sqrt{A_{\text{ref}}}),
\label{eq:new_modele_pression}
\end{equation}
where the {$\beta$} parameter may encompass a vector of parameters describing the mechanical properties of the membrane, related to the distensibility of the vessel: $\beta = 2/\mathcal{D}\sqrt{A} $. 
Causin {\it et al.}[\citep{causin2017mathematical}] have identified two cases:
\begin{enumerate}
\item[--] Thick-walled case: vessel wall thickness is significant, e.g. $h \sim 35\% D$, then:
\begin{eqnarray}
p_{\text{ref}} & = & \frac{2(R_{\text{ref}} + h_{\text{ref}} )^2}{(1-\nu)R_{\text{ref}}^2 + (1+\nu)(R_{\text{ref}} + h_{\text{ref}} )^2}\,p_{\text{ext}}, \nonumber \\
\beta & = & \frac{E((R_{\text{ref}} + h_{\text{ref}} )^2-R_{\text{ref}}^2)}{(1-\nu)R_{\text{ref}}^2 + (1+\nu)(R_{\text{ref}} + h_{\text{ref}} )^2}\cdot\frac{1}{R_{\text{ref}}\sqrt{\pi}}.
\label{eq:wall}
\end{eqnarray}
Geometries measured {\it in vivo} $(R_{\text{d}},h_{\text{d}})$ do not correspond to unloaded conditions. In practice, an inverse problem has to be solved, whose unknowns are $(R_{\text{ref}},h_{\text{ref}},p_d)$. 
As we do not rely on medical imaging and measurements, we assume without loss of generality that the measured configuration corresponds to zero transmural pressure, i.e. $A_{\text{d}} = A_{\Delta p_t=0}$, $h_{\text{d}}=h_{\Delta p_t=0}$. Under this assumption, the problem may be inverted:
\begin{equation}
R_{\text{ref}} = \frac{R_{\Delta p_t=0}}{1-\frac{1-\nu}{E}p_{\text{ext}}}, \quad
h_{\text{ref}} = \sqrt{R_{\text{ref}}^2 +h_{\Delta p_t=0}(h_{\Delta p_t=0}+2R_{\Delta p_t=0})} -R_{\text{ref}}.
\label{eq:h_undeformed}
\end{equation}
\item[--] Thin-walled case: i.e. $h < 10\% D$, which often leads to the following implicit simplification: $h \ll D$. Expressions in Equation \ref{eq:wall} simplify to: $p_{\text{ref}} = p_{\text{ext}}$ so the transmural pressure is null $\Delta p_{t}=0$, the reference surface becomes $A_{\text{ref}} = A_{\Delta p_t=0}$, $h_{\text{ref}}=h_{\Delta p_t=0}$ and $\beta = \frac{E\,h_{\Delta p_t=0}}{(1-\nu^2)\sqrt{\pi}\,R_{\Delta p_t=0}^2}$.
\end{enumerate}
In {Figure~\ref{fig:generic_configuration2}(b)}, we show examples of transmural pressures obtained from the two different models. 
In the simulations, structural model will be selected based on vessel wall thickness, cf. table (\ref{tab:vascular_structure}).

\subsection{Numerical approximation method}
\label{subsec:num}
Getting back to our original ROM describing a {\it single} vessel in Equation \ref{eq:govEq}, 
we now write it under its quasi-linear form:
\begin{equation}
\frac{\partial\vecU}{\partial t} + \matH(\vecU)\frac{\partial\vecU}{\partial x} + \vecS(\vecU)=0,
\label{eq:noncons}
\end{equation}
with,
\begin{equation}
\vecU = \begin{pmatrix} A \\ u \end{pmatrix}, \quad 
\matH = \frac{\partial\vecF}{\partial\vecU} = \ppmatrix{u & A \\ \frac{1}{\rho}\frac{\partial \psi}{\partial A}+(\alpha-1)\frac{u^2}{A} & (2\alpha-1)u} \quad \text{and} \;\; \vecS(\vecU) = \begin{pmatrix} 0 \\ \frac{f}{\rho A} \end{pmatrix}. \nonumber
\end{equation}
Because $A \geq 0$, the Jacobian matrix $\matH$ possesses two real and distinct eigenvalues of different signs: $\lambda_{1,2}=\alpha u \pm \sqrt{c^2+\alpha u^2(\alpha-1)}$, with $c=\big ( \frac{A}{\rho}\frac{\partial \psi}{\partial A} \big ) ^{1/2}$ the pulse wave speed, and {Equation \ref{eq:noncons}} is strictly hyperbolic. It is then possible to diagonalize the system and perform a characteristics analysis, which is very useful to treat the Riemann problems involved at vessels interface and boundary conditions.[\cite{QUARTERONI2016193}]
In Section (\ref{subsec:MCmodel}), we have briefly motivated the fact that the source term will have an effect of varying {stiffness} depending on the flow regime. 
Inspired by earlier approaches,[\cite{pan2014Pulsatile,lee2007}] {authors of the present document} have decided to capitalize on the numerical solver documented in our previous studies and inherited for the case of convection-dominated macroscale arteries.
[\cite{dumas2016,brault2017}] Instead of domain-decomposing the network, in order to select most efficient numerical schemes depending on flow regimes, we have instead treated the network as a whole and computationally improved our solver to handle the various blood flow scales of our system. As explained later, much efforts have been dedicated to  time-stepping tuning and memory constraints originating from the microcirculation. 

\subsubsection{Discontinuous Galerkin scheme}
\label{subsubsec:scheme}
We adopt a DG method with a spectral$/hp$ spatial discretization with $N_e$ non-overlapping regions in each vessel and Legendre $L_{j=0,\dots, p}(x)$ polynomials, {which} is a very efficient framework for high-order discretization of convection-dominated flows, with low dispersion and diffusion errors. The approximated solution is noted: $\vecUd$ in the network domain $\Omega$, and $\vecUd_{\Omega_e^n}$ refers to its components in a particular cell $(e)$ of one of its arterial segment $(n)$. 
The discretized problem 
to solve forms of a system of ordinary differential equations for the modal coefficients of $\vecU$:
\begin{equation}
\label{eq:shemaDG3}
\frac{d \hat{\vecU}^j_{i,\Omega_e^n}}{dt} = \mathcal F(\vecUd_{\Omega_e^n}), \;\; j=0,\ldots,p_e, \;\; i=1,2, \;\; e=1,\ldots,N_e, \;\text{and}\; n=1,\ldots,N,
\end{equation}
where the right hand-side is made of several contributions:
\begin{equation}
\mathcal F(\vecUd_{\Omega_e^n}) = - \frac{1}{J_{\Omega_e^n}} \left( \frac{\partial\vecF_i(\vecU^{\delta}_{\Omega_e^n})}{\partial x}, L_j  \right )_{\Omega_e^n} - \frac{1}{J_{\Omega_e^n}} \left[L_j \left[ \vecF^u_i - \vecF_i(\vecU^{\delta}_e) \right] \right]_{x_{\Omega_e^n}^{-}}^{x_{\Omega_e^n}^{+}} +\left (\vecS_i(\vecU^{\delta}_{\Omega_e^n}),L_j \right)_{\Omega_e^n},
\label{eq:RHS}
\end{equation}
with inner products evaluated by numerical integration, thanks to Legendre-Gauss-Lobatto quadratures. An accurate calculation of the upwinded fluxes $\vecF^u_i$ across inter-elemental boundaries is required to solve the Riemann problem arising at each element interfaces. This is achieved by an efficient Riemann solver of Roe.[\cite{BOLLACHE2014424}]
This  well documented scheme has proven its efficiency for simulating hemodynamics in large- and medium-sized pulsatile arteries.[\cite{QUARTERONI2016193}] 

\subsubsection{Boundary conditions}
The system requires a single inflow boundary condition (BC), and numerous outflow terminal BCs  which are critical to the blood perfusion. A canonical time-dependent total volume flow rate is prescribed at the entrance of the network root in the form of: a half-sine function spanned over the systolic time (1/3 of the cardiac cycle) and null over diastolic time. At the exit of peripheral vessels, linear lumped parameter (0D) models are coupled to the global ROM{, cf. for details about standard lumped parameter outflow models [\cite{Alastruey_2008}]}.
Matched three-element Windkessel models are used for small arteries and arterioles of size $D \geq 100\mu m$ while single-resistance models are used for smaller vessels and capillaries where fluid resistance dominates over wall compliance and fluid inertia.[\cite{Lorthois2011}] 
While other BC closures are possible,[\cite{olufsen2000numerical,Perdikaris2015}] we have estimated 0D BC parameters, i.e. resistance and compliance (when needed), from our network scaling laws (cf. table \ref{tab:random_growth}), under  assumption of Poiseuille law and taking into account the effective non-Newtonian viscosity. This approach resulted in peripheral resistance and compliance distributions in agreement with the literature.
For instance, we have recovered the 
exponential growth of the peripheral resistance with the increasing number of generations of bifurcations.[\cite{phdArimon}] Single-resistance models attached to capillaries exit are tuned to obtain an average pressure drop of about $5\, mmHg$ over the tube.[\cite{Possenti2019}]
In the three-vessel splitting flow junction, encountered many times in the system, the type of
interface condition presented by Quarteroni et al.[\cite{QUARTERONI2016193}]
which is based on consideration of the energy
inequality in the vascular system, was chosen. We therefore impose mass and total pressure conservation.
We mention that the current junction model does not account for momentum loss due to red cells and vessels deformation, bifurcation angles, mechanical properties, etc... but this approximation is probably less important in the microcirculation regime because of the dominance of the viscous forces and the resulting slower flow velocities.

\label{subsubsec:BC}

\section{Numerical simulations on an {idealized} human multi-scale vascular network}
\label{sec:results}

The new implementation of the numerical scheme allows dynamical simulations of large compliant networks.  
We have simulated fifteen cardiac cycles in a compliant vasculature ranging from macro to mesoscale differing by more than two orders of magnitude, cf. Figure~\ref{fig:network_small}.
The choice of the different physical and numerical parameters are summarized in table (\ref{table-simulation-parameter}). {At the inlet, we have imposed an idealized half-sine input flow wave of period $t_c^{\text{sys}}=0.3s$, corresponding to the systolic ejection, for a total cardiac cycle of 1s duration. } Two simulations were compared: one with constant blood viscosity and another one accounting for non-Newtonian effects through the dynamic evaluation of the effective viscosity, {one of the objectives being to quantify some differences between the two time-dependent simulations. }

\begin{table}
    \caption{Main physical and numerical simulation parameters.}
    \label{table-simulation-parameter}
    \centering
    \begin{tabular}{l l}
        {Root vessel diameter $[mm]$} & {$D_{\text{root}}=1$} \\
        {Minimal capillary diameter $[\mu m]$} & {$D_{\text{min}}=8.5$} \\ 
        Blood density $[{kg\cdot m^{-3}}]$: &  $\rho=1050$ \\
        Plasma kinetic viscosity $[m^2\cdot s^{-1}]$: & $\eta = 10^{-6}$ \\
        Discharge hematocrit:  & $H_d = 0.45$ \\
        Cross-section velocity profile: & $\alpha = 4/3$ {(parabolic)} \\
        Membrane Poisson ratio:  &  $\nu = 1/2$ \\
        Membrane Young modulus $E$: & {piecewise linearly} interpolated from [\cite{causin2017mathematical}]\\
        Venous pressure $[mmHg]$: & $ p_{\text{venous}} = 10 $ \\
        Interstitial pressure $[mmHg]$: & $ p_{\text{ext}} = 10 $ \\
        Mesh cells per vessel: & $e=1$  \\
        Legendre polynomial degree: & $p=7$ \\
        Over-integration factor:  & $3/2$ \\
        Number Gauss-Legendre quadrature points/cell: & $n_q=11$ \\
        Time-step $[\mu s]:$  &  $\Delta t =0.3$ \\
        Simulation duration $[s]$: & $T_{\text{final}}=15$ \\
          Inlet BC (pulsatile {half} sinusoidal flow wave) $[ml\cdot min^{-1}]$: 
          & $\bar{Q}_{inlet}=1.74$
        \\
        Systolic duration $[s]$: & $t_c^{\text{sys}}=0.3$ \\
        {Pulse pressure frequency $[Hz]$}: & {$f=3.33$} \\
        Heart beat frequency $[bpm]$: &  $f_{HB} = 60 $ \\
        Outlet BC (3-element Windkessel-type: RCR) & for $D_n \geq 100 \mu m$ \\
        Outlet BC (1-element Windkessel-type: R)   &  for $D_n < 100 \mu m$ \\
    \end{tabular}
\end{table}

\subsection{Blood flow distribution}
\label{subsubsec:flow}

\begin{figure}[ht]
\begin{center}
\includegraphics[width=10cm]{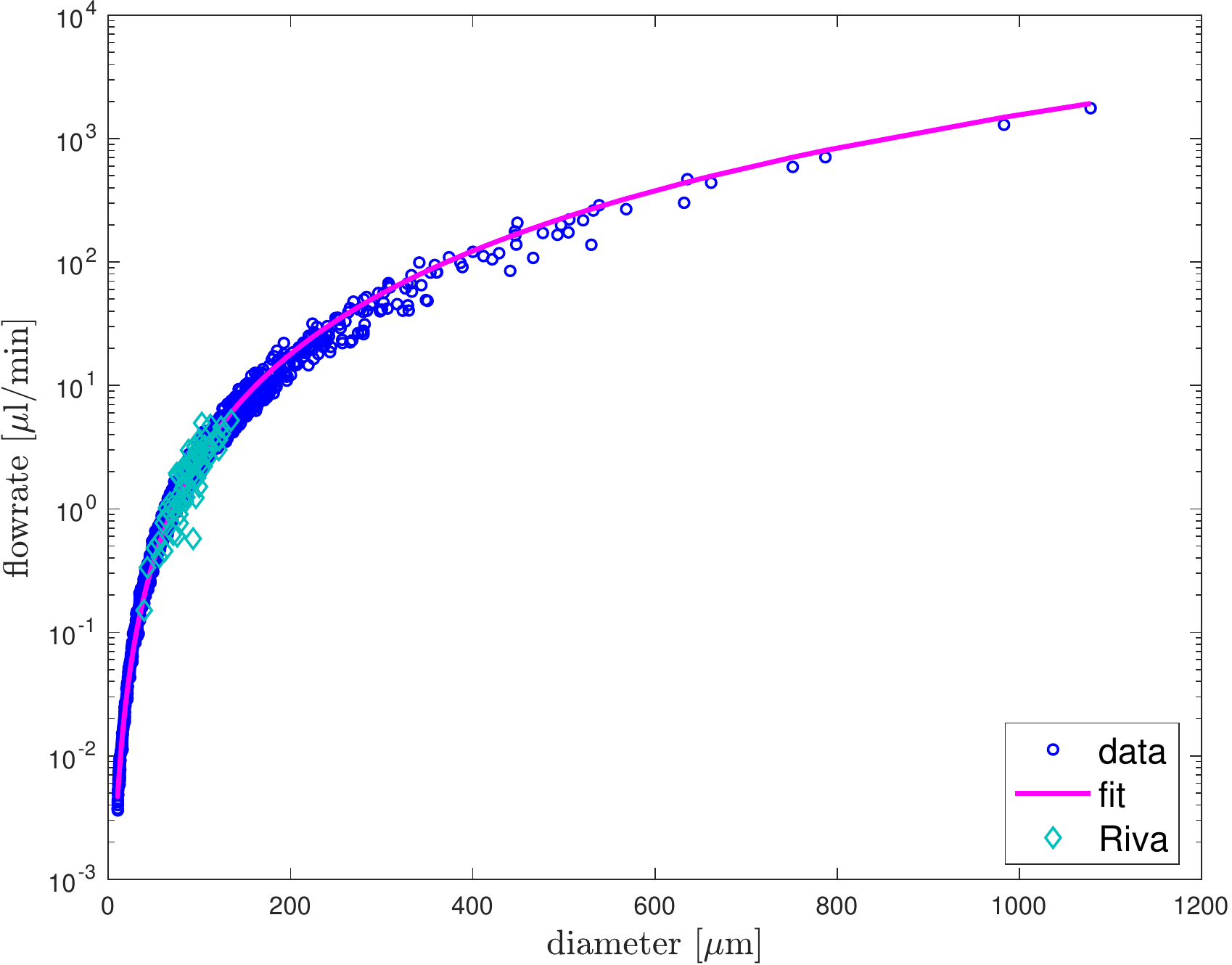}
\caption{Mean volumetric flow rate $Q$ ({represented by blue circles} integrated over each vessel) vs vessel diameter $D$. Continuous line is the power curve fit (with coefficient of determination $r^2=0.99$) to the data points: $Q$ varies as $D^{2.77}$, in very good agreement with the experimental work of Riva {\it et al} [\cite{Riva1985}] (cf. $Q \sim D^{2.76}$, {cf. Riva's data represented by cyan markers with diamond shapes}) and close to the cube law described by Zamir.[\cite{Zamir2000}]}
\label{fig:debitvsdiam}
\end{center}
\end{figure}

The entire tree was well perfused according to the mean volumetric blood fluxes observation, cf. Figure~\ref{fig:debitvsdiam}. The simulation generated an unsteady flow rate with an average of  $1.74\,ml\cdot min^{-1}$ at the main feeding artery and  $5\,nl\cdot min^{-1}$ in capillaries. The agreement between the data fit and experimental measurements in humans is excellent. [\cite{Riva1985}] The decline of the corresponding averaged blood flow velocities (not represented) at three different scales along the arteriolar tree are also in good agreement with the literature, e.g.  [\cite{Van2Vosse_LN,Possenti2019}]: for -- small arteries at $D\sim 600\,\mu m$: $u_{\text{sa}}\approx 25\, mm\cdot s^{-1}$, -- arterioles at $D\sim 40\,\mu m$: $u_{\text{art}}\approx 3\, mm\cdot s^{-1}$ and -- capillaries at $D\sim 10\,\mu m$: $u_{\text{cap}}\approx 0.85-1\, mm\cdot s^{-1}$, 
respectively.

\subsection{Pressure and pulsatility}
\label{subsubsec:pulsatility}

\begin{figure}[ht]
\begin{center}
\includegraphics[trim=2.5cm 0cm 0cm 0cm,clip,width=7.5cm]{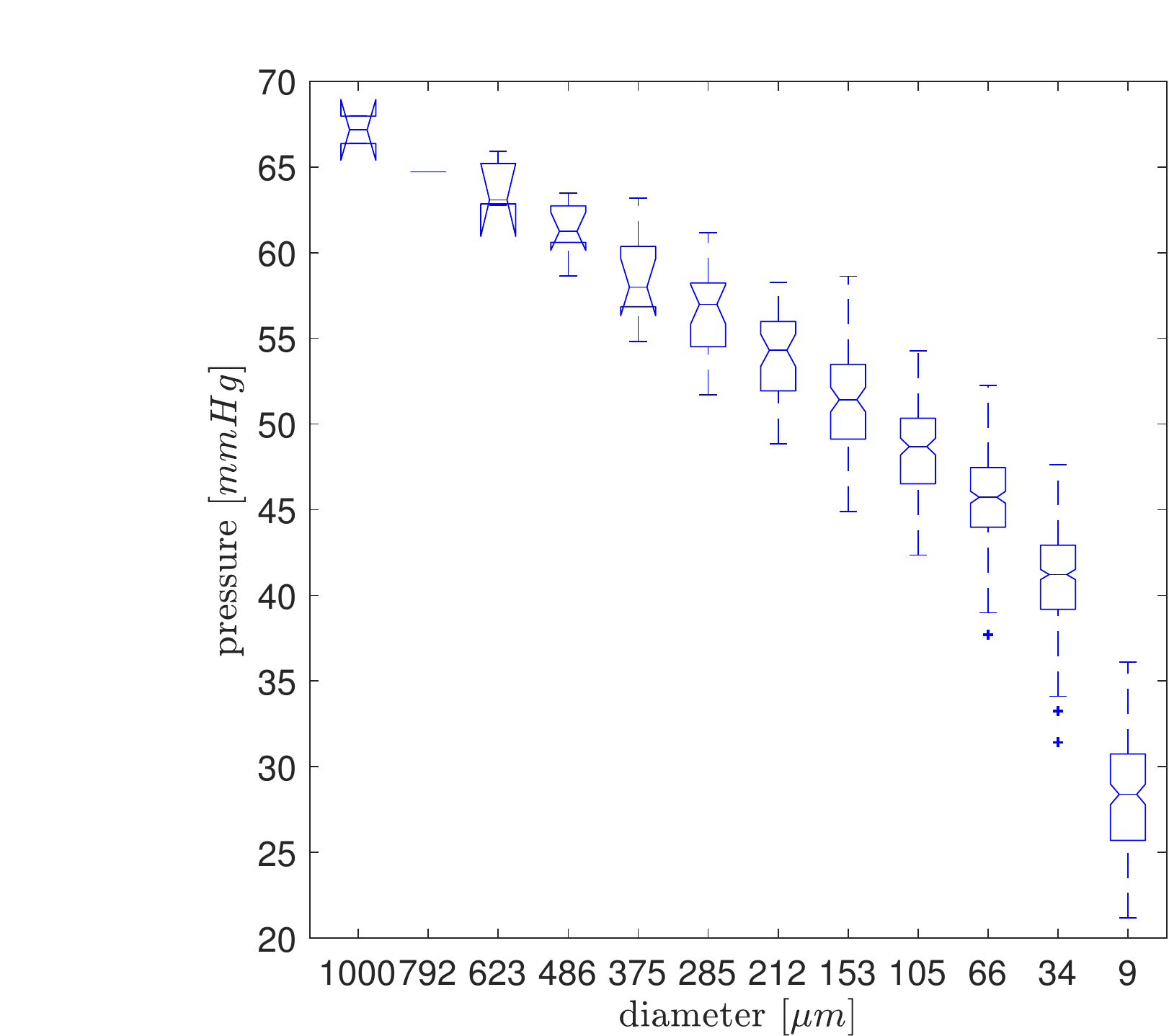}(a)\includegraphics[trim=8.5cm 0cm 0cm 0cm,clip,width=7.53cm]{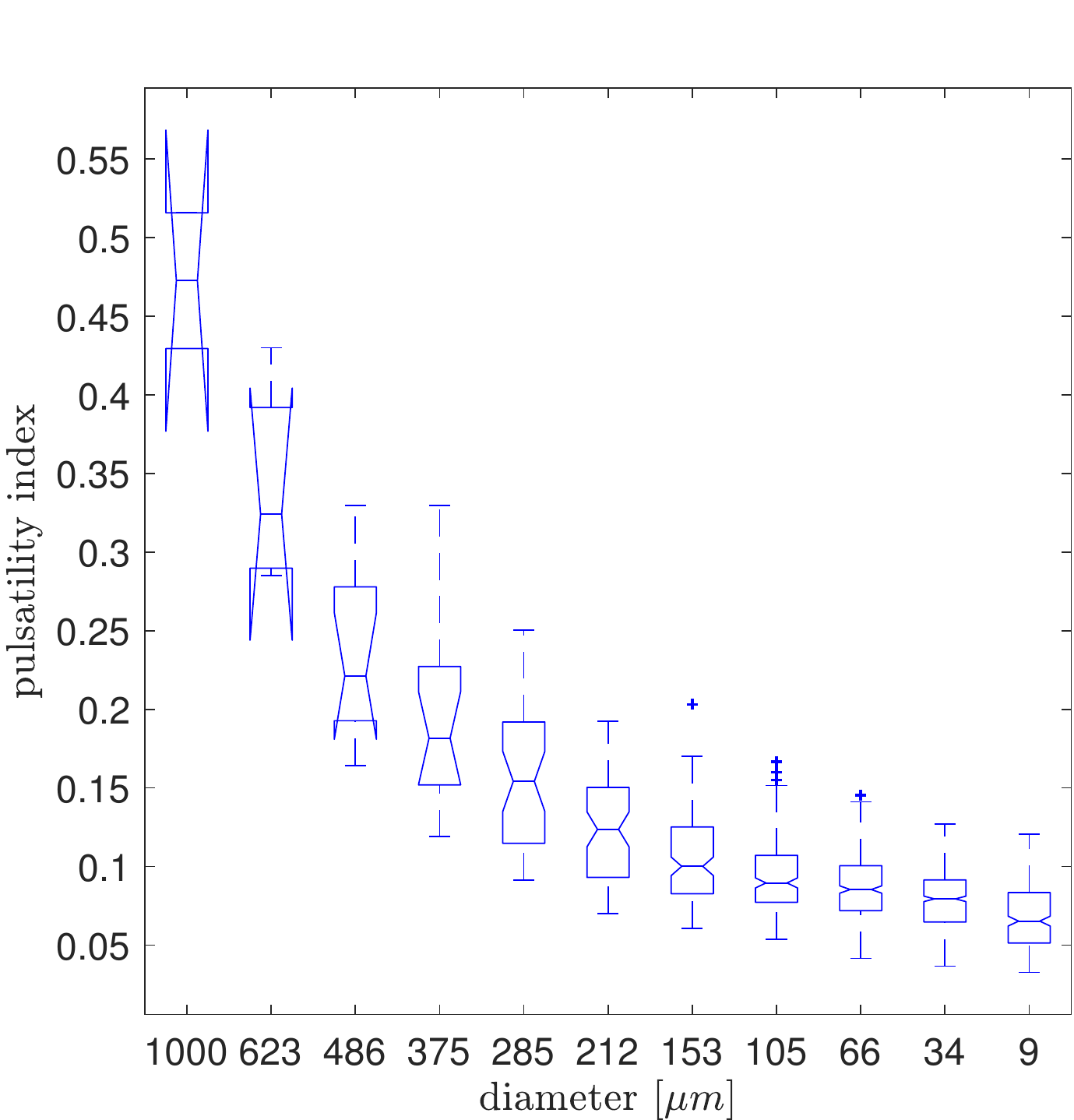}(b)
\caption{Blood pressure-related notched Box-plot type statistics vs. diameter size. (a): mean pressure and (b): pressure pulsatility index. On each box, the {horizontal} central mark indicates the median value, {and the notch outlines a $95\%$ confidence interval around this  value. For some insufficient sample sizes, especially at large scale, the confidence interval is sometimes overestimated which explains the buckled box shapes}.
The bottom and top edges of the box represent the 25$^{\text{th}}$ and 75$^{\text{th}}$ percentiles, respectively. The whiskers extend to the most extreme data points not considered outliers (plotted individually using the $(+)$ symbol).}
\label{fig:bplot_pulse}
\end{center}
\end{figure}

The boxplot statistics of averaged (over space-time) pressure signals, plotted against vessel diameters, cf. Figure~\ref{fig:bplot_pulse}(a), show a strong pressure drop from the main feeding artery to the capillaries. {As expected,} mean pressure variability is largest {across the arteriolar span, i.e. $100 \mu m \leq  D \leq 10\mu m$}. Newtonian results (not presented) exhibit a stiffer pressure drop with significant prediction {different} far upstream in the network, as observed in {Perdikaris \it et al.[\cite{Perdikaris2015}]}. 
The dynamic characteristics of the pressure signals, captured by the pressure pulsatility index (here measured at the center of each vessel): $p_I=(p_{\text{systolic}}-p_{\text{diastolic}})/p_{\text{mean}}$, cf. {Figure~\ref{fig:bplot_pulse}(b)}   also present strong decay with one order of magnitude difference from central to most distal regions. Nonetheless, pulsatility index in capillaries region do remain significant. Interestingly, statistical linear regressions of the first two $p_I$ moments against diameters (not included here) reveal a close-to-linear behavior of its mean ($r^2=0.98$; $p{\text{-value}}=1.8\cdot 10^{-9}$) and its standard deviation distribution ($r^2=0.96$; $p{\text{-value}}=7.8\cdot 10^{-7}$), respectively.

\subsection{Wall shear stress and pulse wave speed}
\label{subsubsec:wss}

\begin{figure}[ht]
\begin{center}
\includegraphics[width=7.5cm]{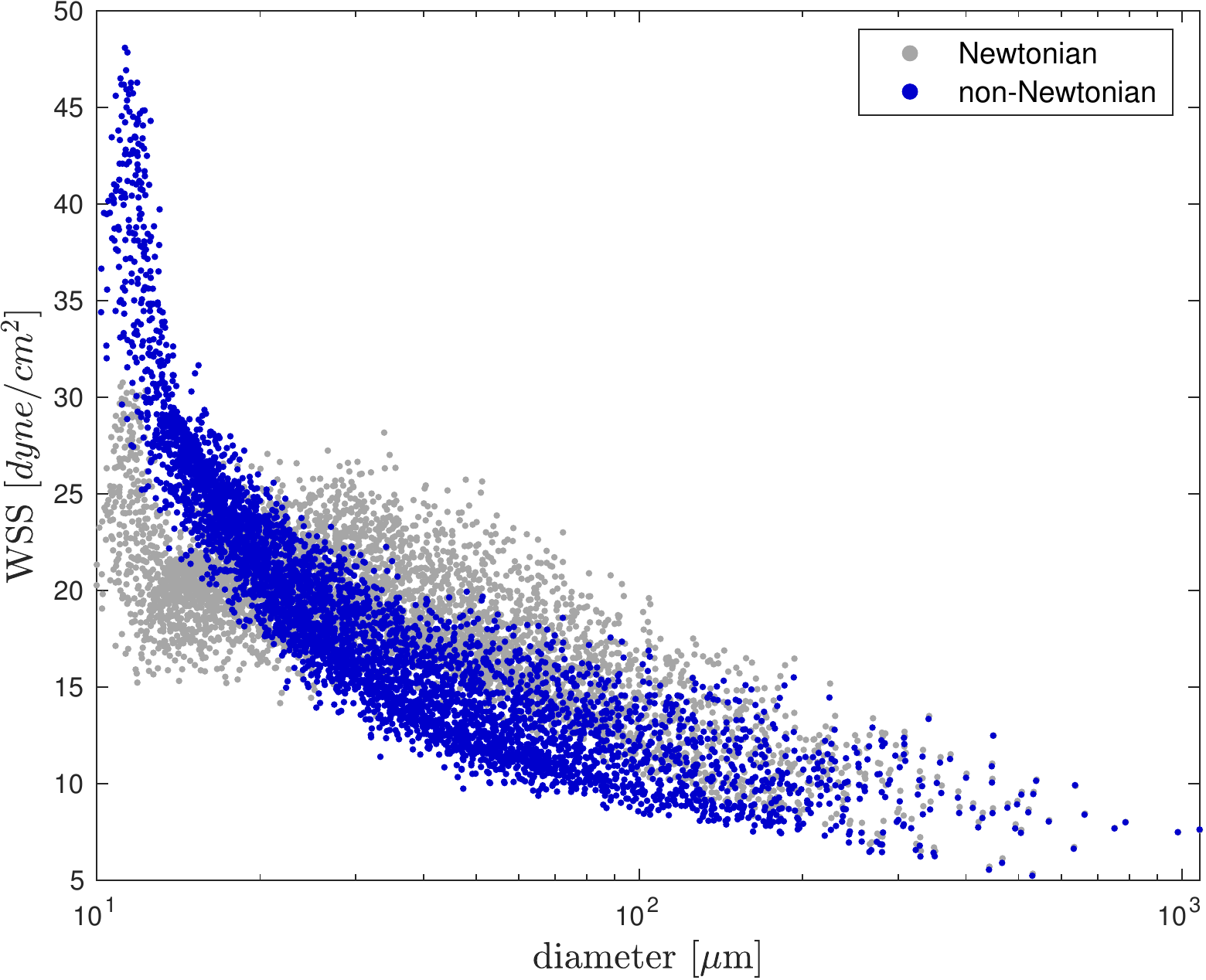}(a)\includegraphics[width=7.5cm]{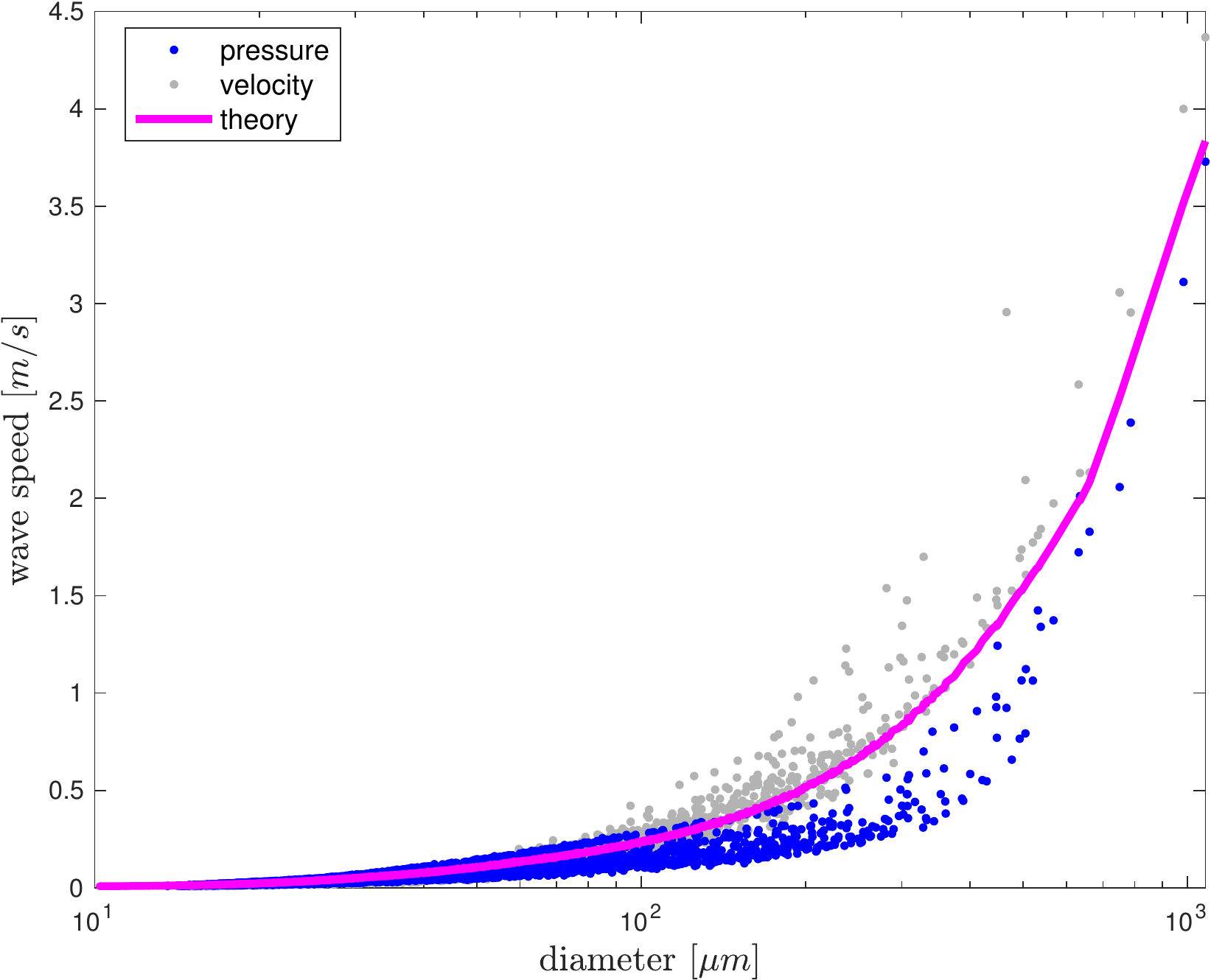}(b)
\caption{(a): mean wall shear stress (WSS) vs. mean lumen diameter. 
(b): mean pressure and velocity wave speeds (PWV) vs. mean lumen diameter; the solid line refers to the theoretical pulse wave speed calculated according to { Equation \ref{eq:wavespeedWomersley}} with respect to local reference lumen diameter and  Womersley number $W_{\alpha}$. For both plots, each dot represents the statistics acquired on a single vessel.}
\label{fig:wss}
\end{center}
\end{figure}

While our  model contains  simple non-Newtonian effects compared to other more complex constitutive models [\cite{GHIGO2018}], it does include temporal and spatial structure-dependent viscous effects. In low shear regions, it is therefore important to quantify these effects. We monitor the wall shear stress (WSS), cf. {Figure~\ref{fig:wss}(a)}, that we relate to the wall shear rate through the effective viscosity $\mu(x,t)$:
\begin{equation}
\tau_{rx}(x,t)\equiv {\textrm{WSS}}=\mu \left ( \frac{\partial \Tilde{u}}{\partial r}\right ) _{r=R}=\mu(x,t)\, u(x,t)\left ( \frac{\partial \phi}{\partial r}\right ) _{r=R} \quad \left ( =4\mu \frac{u}{R} \text{ for Poiseuille profile} \right ).
\label{eq:WSS}
\end{equation}
Only mean values over space/time are reported.
Blue dots refer to the WSS obtained from the simulation with non-Newtonian rheology. As expected, despite a wide spread, WSS increase hyperbolically as arteriolar diameter decreases. Moreover, it is observed that WSS values are in good agreement with {\it in vivo} measures from recent literature.[\cite{Koutsiaris2016}] Grey dots depict the Newtonian WSS response. 
They present values obtained from the simulation with {\it constant} blood viscosity ($\mu_{\text{constant}}=3.4 \,Pa \cdot s$). The behavior is completely different: WSS increase stagnates for diameters lower than $50\mu m$. Non-Newtonian effects are therefore significant in most of the domain (not in small arteries) and must imperatively be accounted for.  

The speed distribution of the pulse waves in the network are also worth investigation. Microvascular pulse wave velocity (PWV) emerges as a novel indicator of hypertension, but its dependence on various vascular scaling properties remains unclear. 
The theoretical PWV is evaluated , {for a pressure pulse of frequency $f=3.33Hz$}, thanks to Equation \ref{eq:wavespeedWomersley}, {under the favorable assumption whereby an antegrade pulse is transmitted free of reflection at branch points,} and compared to the mean simulated wave speeds, cf. {Figure~\ref{fig:wss}(b)} where only internal network vessels results are depicted. PWVs
are directly processed from the time-dependent pressure and velocity wave signals. The PWV range is very wide between fast entering waves ($\sim 4 m/s$) to slow travelling waves ($\mathcal{O}(cm/s)$) in the capillaries. {The results agree well with the available literature. For instance, values obtained from the pressure signals in arterioles $D\approx 200 \mu m$ are within a $[11-13cm/s]$ range compared to $[7-11cm/s]$ obtained by \citer{Intaglietta et al.} [\cite{Intaglietta1971}] for the cat omemtum at slightly lower frequency $f=2.5Hz$, phase velocities increasing with frequency [\cite{Salotto1986}]}.

\section{Discussion} 
\label{sec:discussion}
\begin{figure}[ht]
\begin{center}
\includegraphics[width=10cm]{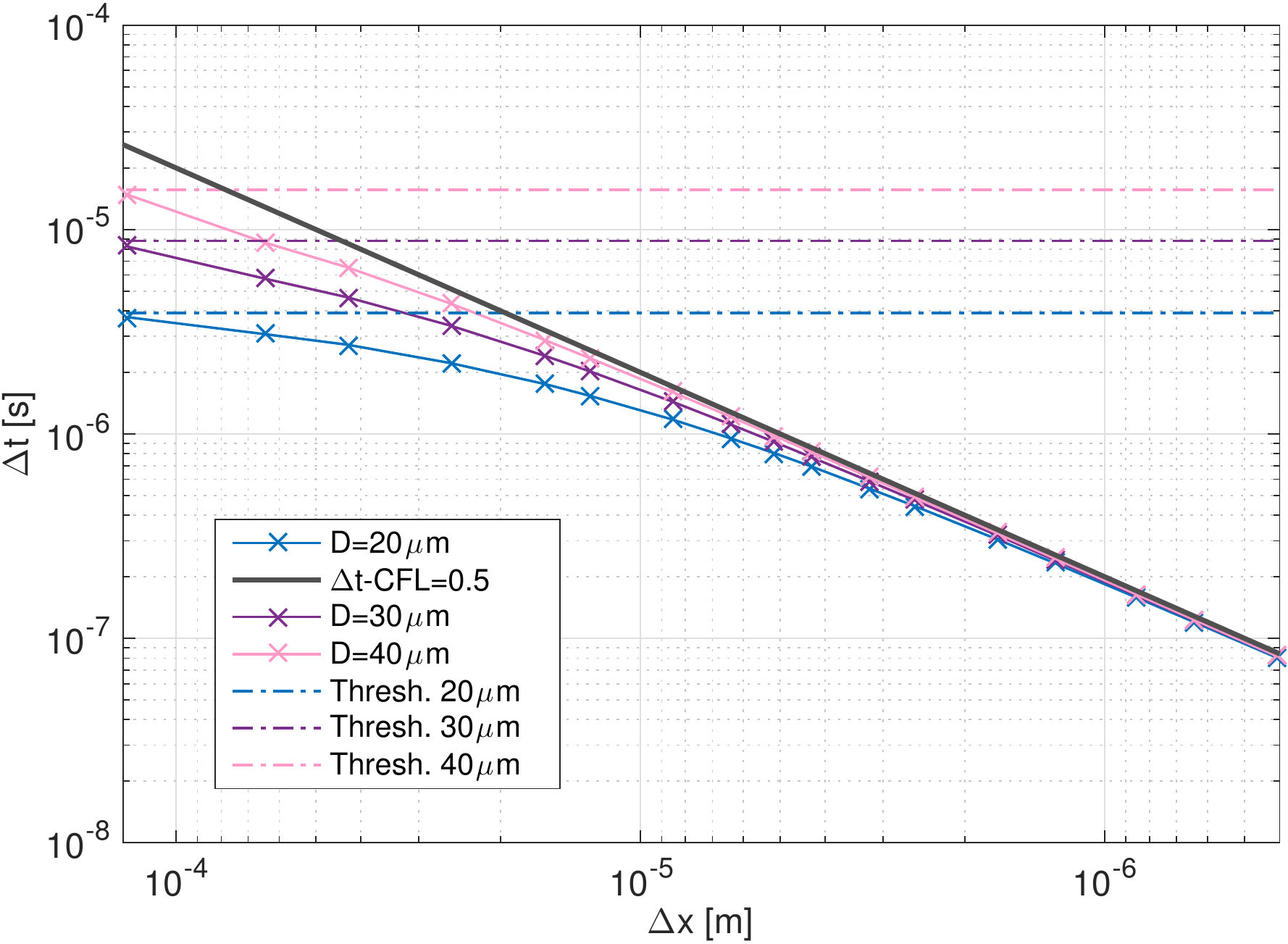}
\caption{Numerical determination of the largest timestep allowing for a stable macro/meso blood flow simulation, given certain axial and radial discretization scales (solid curves with crosses). The oblique black solid line refers to the macroscale $\Delta t^{\text{CFL}} $ barrier, while horizontal dotted-dashed lines refer to the mesoscale limitation imposed by the diffusion time related to the radial coordinate. Three different diameter sizes (in $\mu m$) are investigated.}
\label{fig:dt_CFL}
\end{center}
\end{figure}

The specifics of proposed modeling are discussed in details. The model bridges a macroscale $\mathcal{O}(cm-mm)$ to mesoscale $\mathcal{O}(\mu m)$ vascular territory and may be used to explore its baseline structure and function relationship through numerical blood flow simulations. Only arteries, arterioles and capillaries are modeled, avoiding the need for complex venular tree modeling.[\cite{causin2017mathematical}] 
Despite spatial dimension simplifications, hemodynamics simulation across several scales is a very demanding computational task. The new implementation makes feasible the {\it dynamical} simulation of large compliant network  with $\mathcal{O}(10^4)$ vessels, without resorting to domain decomposition, 
thanks to the rapid convergence properties exhibited by the spectral elements method, with reasonable dimensionality: spatial number of degrees of freedom: $2\times N \times e \times \lceil \frac{3p}{2} \rceil \sim \mathcal{O}(10^5)$. 
Computer time to complete one cardiac cycle (with $\Delta t=0.3\mu s$) was 5 hours for an Intel Xeon E5 CPU having 3.7GHz clock frequency and 16Gb DDR3 of RAM. In case of larger systems, DG methods compactness (elements are discontinuous and inter-element communication is minimal) make them great candidates for efficient parallelization.[\cite{Perdikaris2015}]
However, (semi-)implicit formulations are preferred for 3D models, the solver relies on an ($2^{\text{nd}}$-order Adams-Bashforth) explicit scheme. It performs well under convective stability constraint of CFL-type , i.e. $\Delta t^{\text{CFL}} \propto \frac{1}{2} \frac{\Delta x}{\lambda_{\text{max}}}$ ($\lambda_{\text{max}}$: maximum eigenvalue of the Jacobian matrix), but for system with large physical or numerical diffusive effects (low Reynolds and Womersley numbers), it is more strongly impacted by the diffusive stability constraint $\Delta t^{\text{diff}} \propto \frac{\Delta R^2}{\nu} $. Therefore, depending on the discretization retained, time-stepping  choice insuring stability is not a straightforward task because of the interplay between these different effects. In Figure~\ref{fig:dt_CFL}, we present an example of optimal time-stepping which is {\it numerically} determined for different discretization sizes. When axial discretization is fine enough, the time-step must asymptotically evolve as $\Delta t^{\text{CFL}} $, while at the opposite, it is fully dictated by the $\Delta t^{\text{diff}} $  scale, which is particularly stringent for vessel of smaller diameter. \\
In order to make the time-advancing computationally tractable, we have  vectorized the computation of {Equation \ref{eq:RHS}}, in the form of a dense matrix structure encapsulating the entire network information.
Other studies have proposed to make implicit the time-integration, 
to the price of additional numerical effort in terms of improvement of system conditioning or sparse algebra acceleration  [\cite{lee2007,pan2014Pulsatile}].
Another option, in the context of partitioned schemes, implemented in sequential or parallel fashion, is to resort to multi-level time stepping techniques,[\cite{Blanco2012}] where the different geometric components are solved with individual convenient time steps and matching interface conditions are fulfilled thanks to suitable synchronization procedures.\\
The model is conveniently compatible with a mixed bag of BCs and wall structural models, allowing fine tuning of physiological effects depending on the vessel scale. Moreover, the large scale representation authorizes a certain resilience to the choice of inaccurate outflow boundary conditions. Indeed, for macroscale systemic circulation, this is usually a difficult and unreliable task based on the calibration of lumped parameters models, that is somewhat mitigated in our case to the price of specifying the mechanical properties of the mesoscale vessels.\\
It is time-dependent and therefore offers great potential to study biomechanical determinants of pulsatility damping in great details through the vasculature. 
Results from {Figure~\ref{fig:wss}(b)} quantify to what amount PWVs are slowed down along the arteriolar region. {Accordingly with linear pulse transmission literature, pressure pulsation persists throughout the microvasculature and pulse propagation speed in small vessels is observed to be orders of magnitude less than predicted by the Moens-Korteweg relation.}
Despite the large variation, {especially at large scale, potentially related to reflections at branching junctions}, the agreement with the theoretical prediction is quite satisfactory [\cite{Salotto1986}]. Interestingly, pressure (velocity) wave speeds are a bit slower (faster) than the reference curve for medium size vessels, respectively. Eventually both quantities reach very comparable PWVs once they get closer to the capillaries, which is coherent with the assumption of synchronization of pressure and flux waves in microcirculation. {More realistic viscoelastic model describing vascular wall properties would be interesting to include in our multiscale model, but should not qualitatively affect the results, despite increasing pulse attenuation and affecting pulse transmission time.  }\\
The model is applicable to large-scale vascular networks. A noticeable effort has been made in order to craft vascular structure with realistic topology.
Some adjustments to the dyadic structure have been done in order to account for the capillaries density, and scale-specific pruning gradation of the tree has been proposed to minimize the sensitivity of the cut-off radius of the tree.
. Nevertheless the effect of the tortuous morphology of the small vessels is neglected by the model. 
The arrangement of capillaries also strongly depends on the host organ and type of tissue. For instance, a network like configuration of the capillary bed, more realistic to model tumoral tissues, would require adding network anastomoses to the geometry (not a blocking point).
Moreover, tumoral vasculature is made of tortuous and dilated vessels which are spatially strongly heterogeneous. In this case, Murray's law does not hold anymore.
But the proposed computational approach is completely general with respect to more complex and accurate large-scale human morphology, obtained from micro-computed tomography imaging for instance.[\cite{Stamatelos2014}]\\
In order to gain reliability and versatility, ROM must propose modeling improvements leading to {\it significant} error prediction reduction. Otherwise, low gains are very difficult to detect below the inherent approximation errors, and demanding solver developments become in this case hard to motivate. 
An experimentally-characterized \farhaeus-Lindqvist effect, inducing a diameter temporal dependence of viscosity, was added to the model and had a major impact on the overall flow, especially at the mesoscale. In order to account for phase separation effects, more involved modeling including an additional conservation equation for the RBCs may be introduced relating $H_d$ and $H_t$ through a time-dependent passive scalar transport equation where $H_d$ is advected at the flow velocity. 
We have implemented this approach but obtained unphysical oscillating hematocrit concentration results.
These oscillations were not only detrimental to the numerical simulation (unstable situation) but they also provided very different results compared to the much smoother observed concentrations for experimental injection of a passive compound in arterial circulatory subsystem.[\cite{Boyer_2015}] \\
While it is known that high-order DG methods without slope limiters dealing with problems governed by hyperbolic conservation laws suffer from spurious oscillations in the vicinity of discontinuities, we believe that more work is needed to understand the limitation of the one-dimensional cross-section averaged modeling for multi-scale transport.
Constant concentration cross-section profiles  are in this case unrealistic. Thus, uneven concentration of agent propagation in the vessel lumen and diffusion effects should be integrated in the ROM.\\
In conclusion, this ROM model may serve as a useful tool for various purposes, such as the investigation of pulsatility dependent microvascular physiology. Initial results have been encouraging and dynamical effects related to the coupling between the non-Newtonian blood rheology and flow pulsatility have been well captured.
It may also serve as a foundation for the investigation of many pathologies
related to microcirculation, which is crucial in the development of tumor-induced cancer but also in its detection and treatment, since it is the main path for the delivery of imaging agents and therapy drugs.
Future works will involve model improvements for a better understanding of the micro-environment of vascularized tumors in relation to the systemic circulation. 
In particular, precise modeling of microvascularization is a major challenge today to strengthen and guide the interpretation of perfusion imaging used to monitor therapeutic effects.

\section*{Conflict of Interest} 
The authors declare that there is no conflict of interest regarding the publication of this article.
 
\section*{Acknowledgment}
This work has been supported by the LabeX LaSIPS (ANR-10-LABX-32) managed by the French National Research Agency under the ``Investissements d'avenir" program (ANR-11-IDEX- 0003-02).

\bibliography{libabbr}

\end{document}